\documentclass[a4paper,12pt]{article}
\pdfoutput=1
\usepackage{amsmath,amssymb}
\usepackage{latexsym}
\usepackage{dsfont}
\usepackage{graphicx}
\usepackage{amsopn}
\usepackage{color}
\usepackage{ulem}
\usepackage[width=1.15\textwidth]{caption}
\makeatletter
\let\oldabstract\abstract
\let\oldendabstract\endabstract
\renewenvironment{abstract}
{%
               {\list{}{\addtolength{\leftmargin}{-2.5em} % change this value to add or remove length to the the default
                        \listparindent 1.5em%
                        \itemindent    \listparindent%
                        \rightmargin   \leftmargin%
                        \parsep        \z@ \@plus\p@}%
                \item\relax}%
               {\endlist}%
\oldabstract}
{\oldendabstract}

\makeatletter
\@addtoreset{equation}{section}
\renewcommand{\theequation}{\thesection.\@arabic\c@equation}
\makeatother
\makeatletter
\renewcommand\appendix{\par%\newpage
  \setcounter{section}{0}%
  \setcounter{subsection}{0}%
  \gdef\thesection{Appendix \@Alph\c@section }
  \renewcommand{\theequation}
  {\Alph{section}.\arabic{equation}}
}
\makeatother

\def \be {\begin{equation}}
\def \ee {\end{equation}}
\def \hs {\hspace}
\def \bea{\begin{eqnarray}}
\def \eea{\end{eqnarray}}
\def \nn {\nonumber}

\def \Tr {{\textrm{Tr}}}
\def \a {\alpha}

\def \g {\gamma}
\def \G {\Gamma}

\def \D {\Delta}

\def \o {\omega}

\def \th {\theta}

\def \t {\tau}
\def \z {\zeta}

\def \p {\partial}

\def \nn {\nonumber}

\title{\textbf{Kinematic Space and Wormholes}}
\author{Jian-dong Zhang$^{1}$\footnote{zhangjd9@mail.sysu.edu.cn} and Bin Chen$^{2,3,4}$\footnote{bchen01@pku.edu.cn}}
\date{}

\begin{document}
\maketitle

\vspace{-10mm}

\begin{center}
{\it
$^{1}$TianQin Research Center for Gravitational Physics, Sun Yat-sen University, \\Zhuhai 519082, Guangdong, P. R. China \\
$^{2}$Department of Physics and State Key Laboratory of Nuclear Physics and Technology, Peking University, Beijing 100871, P.~R.~China\\
\vspace{2mm}
$^{3}$Collaborative Innovation Center of Quantum Matter, 5 Yiheyuan Rd, \\Beijing 100871, P.~R.~China\\
$^{4}$Center for High Energy Physics, Peking University, 5 Yiheyuan Rd, \\Beijing 100871, P.~R.~China
}
\vspace{10mm}
\end{center}

\begin{abstract}
The kinematic space could play a key role in constructing the bulk geometry from dual CFT.
In this paper, we study the kinematic space from geometric points of view, without resorting to differential entropy.
We find that the kinematic space could be intrinsically defined in the embedding space.
For each oriented geodesic in the Poincar\'e disk, there is a corresponding point in the kinematic space.
This point is the tip of the causal diamond of the disk whose intersection with the Poincar\'e disk determines the geodesic.
In this geometric construction, the causal structure in the kinematic space can be seen clearly.
Moreover, we  find that every transformation in the $SL(2,\mathbb{R})$ leads to a geodesic in the kinematic space.
In particular, for a hyperbolic transformation defining a BTZ black hole, it is a timelike geodesic in the kinematic space.
We show that the horizon length of the static BTZ black hole could be computed
by the geodesic length of corresponding points in the kinematic space.
Furthermore, we discuss  the fundamental regions in the kinematic space for the BTZ blackhole and multi-boundary wormholes.
\end{abstract}
\baselineskip 16pt
\thispagestyle{empty}

\newpage

\section{Introduction}

One of recent developments in the AdS/CFT correspondence is on the emergence of spacetime and diffeomorphism.
The key notion in the study of the emergent spacetime is the entanglement and its holographic computation.
The holographic entanglement entropy in the Einstein gravity is proposed in \cite{Ryu:2006bv,Ryu:2006ef} to be
\be
S_{RT}=\frac{A}{4G_N},\label{RTform}
\ee
where $A$ is the area of the minimal surface which is homologous to the boundary region.
This formula, being reminiscent of the Bekenstein-Hawking formula for the black hole entropy\cite{Lewkowycz:2013nqa},
suggest a deep relation between quantum gravity and quantum information.
It has been widely suspected that the holographic entanglement entropy
could play a pivotal role in constructing bulk spacetime and even bulk physics.

There are several proposals to construct the bulk geometry from boundary CFT,
mainly based on the concept of the tensor network \cite{Swingle:2009bg,VanRaamsdonk:2010pw,Pastawski:2015qua,Czech:2015kbp,Czech:2015xna,Hayden:2016cfa,Bhattacharyya:2016hbx}.
Among them, one promising approach proposed by B. Czech et.al is to view the MERA(Multi-scale Entanglement Renormalization Ansantz)
tensor network as a discrete version of vacuum kinematic space\cite{Czech:2015kbp,Czech:2015xna}.
This proposal is inspired by the study of the hole entropy in the bulk from dual CFT data,
which suggests a way to define the bulk geometry from differential entropy\cite{Balasubramanian:2013rqa}.
To compute the length of a curve $\g$ in the hyperbolic plane,
one could apply integral geometry rather than differential geometry.
The length could be given by the Crofton formula
\be
\mbox{Length of the curve $\g$}=\frac{1}{4}\int_K\o(\th,\a)n_\g(\th,\a),\label{crofton}
\ee
where $\th$ and $\a$ label the oriented geodesic in the Poincar\'e disk,
$n_\g(\th,\a)$ is the intersection number of the geodesic with the curve $\g$ and $K$ denotes the kinematic space.
The most interesting part is on the measure $\o(\th,\a)$ in the kinematic space, which has the form as
\be
\o(\th,\a)=-\frac{1}{\sin^2\a}d\a\wedge d\th,\label{kinemeasure}
\ee
or in terms of the  coordinates of the ending points of the geodesics  on the disk boundary
\be
u=\th-\a, \hs{3ex}v=\th+\a,\label{endcoord}
\ee
the form of the measure becomes
\be
\o(u,v)=\frac{1}{2\sin^2\left(\frac{v-u}{2}\right)}du\wedge dv.\label{endkinemeasure}
\ee
This measure is more suggestive when being given by
\be
\o(u,v)=\frac{\p^2S(u,v)}{\p u\p v}du\wedge dv,\label{entropymeasure}
\ee
where $S(u,v)$ is the entanglement entropy of the interval $(u,v)$.
In \cite{Czech:2015qta}, the authors furthermore suggest that the Crofton form should be interpretated as the conditional
mutual information\footnote{For the higher dimensional study of the Crofton form and its interpretation, see \cite{Gil:thesis,Huang:2015xca}. }.
The basic picture on the kinematic space is that it is an auxiliary Lorentzian geometry,
whose metric is defined in terms of conditional mutual information.

In this paper, we would like to  study the
kinematic space from geometric points of view
\footnote{While we are preparing this manuscript, there appeared two works \cite{Asplund:2016koz, deBoer:2016pqk},
which partially overlap our discussion in section 2. }.
We show that the kinematic space can be defined in a geometric way.
Simply speaking, every geodesics in the Poincar\'e disk could define a causal cone, whose tips are in the kinematic space.
The causal structure in the kinematic space can be seen clearly in this geometric picture.
Moreover, we discuss the static wormhole solution in the AdS$_3$ gravity and its representation in the kinematic space.
We show that the timelike geodesic in the kinematic space is closely related to the isometric transformation of hyperbolic type.
For the BTZ spacetime formed by the identification of the geodesics with respect to a hyperbolic element in the Fuchsian group,
its horizon length could be read from the timelike geodesic distance in the kinematic space
between the points corresponding to the geodesics in the disk.
Therefore for the eternal BTZ black hole formed by the identification of a pair of geodesics,
it could be described by two timelike separated points in the kinematic space.
These two points cannot be determined uniquely.
As long as a pair of points lie in the timelike geodesic determined by the transformation and the geodesic distance between them is fixed,
they describe the same BTZ spacetime.
On the other hand, the timelike geodesic defined by a hyperbolic transformation is unique.
In this sense, the BTZ spacetime could be related to a timelike geodesic in the kinematic space.
In the similar spirit, we can describe the multi-boundary wormhole easily.

Another interesting issue is to consider the kinematic space for the BTZ wormhole and other multi-boundary wormhole background.
The kinematic space can still be defined by the geodesics in these spacetime.
We  start from the kinematic space for AdS$_3$, and take into account of  the quotient identification defining the wormhole.
We discuss carefully how to classify the geodesics in the BTZ spacetime
and propose a consistent rule to define the fundamental region in the kinematic space for the BTZ spacetime.
We furthermore show that the fundamental region for the multi-boundary wormhole
could be defined to be the intersection of the fundamental regions for the BTZ spacetimes,
each being defined by the fundamental elements of the Fuchsian group.

The remaining part of this article is organized as follows.
In section 2, after giving a brief review of AdS$_3$ spacetime and its different coordinate systems,
we show how to describe the kinematic space.
In section 3, we review the construction of the static BTZ black hole
and general multi-boundary wormholes by using the Fuchsian group identification.
Especially we discuss the three-boundary wormhole and single-boundary torus wormhole.
In section 4, we discuss the properties of the kinematic space.
We show that a $SL(2,R)$ transformation, being the isometric transformation of AdS$_3$,
define  a geodesic in the kinematic space.
In particular, we study the three boundary wormhole to get its fundamental regions in the kinematic space,
and give a method to get the fundamental region in kinematic space for general wormholes.
We end with conclusions and discussions in section 5.

\section{AdS$_3$ and its Kinematic Space}

The AdS$_3$ can be taken to be a hyperboloid space in the $2+2$ dimensional flat spacetime $\mathbb{R}^{2,2}$ with the metric
\be
ds^2=-dU^2-dV^2+dX^2+dY^2,\label{4Dflat}
\ee
The AdS$_3$ spacetime is defined by the relation
\be
-U^2-V^2+X^2+Y^2=-l^2,\label{AdSembed}
\ee
where \(l\) is the AdS radius.
For simplicity, we set \(l=1\) in this paper.
Defining the coordinates
\bea
&U=\cosh\rho\cos\tau,~~~&V=\cosh\rho\sin\tau,\nn\\
&X=\sinh\rho\cos\theta,~~~&Y=\sinh\rho\sin\theta,\label{AdSco}
\eea
we can read the metric of AdS$_3$ in the global AdS coordinates
\bea
ds^2=-\cosh^2\rho d\tau^2+d\rho^2+\sinh^2\rho d\theta^2,\nn\\
\tau\in\mathbb{R},~~~\rho>0,~~~\theta\sim\theta+2\pi.\label{AdS}
\eea

The classical solutions in the AdS$_3$ gravity could be constructed by the quotient identification
by the discrete subgroup of the isometry group $SL(2,\mathbb{R})$.
If we focus on the static solutions,
the construction could be understood as the identification of the geodesics pairwise in the constant time slice of AdS$_3$.
The constant time slice is a two-dimensional hyperboloid $\mathbb{H}_2$, the so-called Poincar\'e upper half plane, which is of the metric
\be
ds^2=d\rho^2+\sinh^2\rho d\theta^2. \label{H2}
\ee
In fact, for simplicity we just take the $\tau=0$ slice, this is equivalent to $V=0$.

\subsection{$\mathbb{H}_2$ and dS$_2$}

The relation between the constant time slice of AdS$_3$ and the kinematic space is most easily seen
by embedding them into the three-dimensional flat spacetime,
which is the $V=0$ slice of $\mathbb{R}^{2,2}$
\be
ds^2=-dU^2+dX^2+dY^2. \label{3Dflat}
\ee
The two-sheeted hyperboloid $\mathbb{H}_2$ is then the $V=0$ slice of AdS$_3$
\be
U^2-X^2-Y^2=1. \label{H2embed}
\ee
With the embedding coordinates
\be
U=\cosh\rho,\hs{3ex}X=\sinh\rho\cos\theta,\hs{3ex}Y=\sinh\rho\sin\theta, \label{H2co}
\ee
we recover the metric (\ref{H2}).
If we make a  projection from the point $(U,X,Y)=(-1,0,0)$,
we can project the upper sheet of the hyperboloid onto the unit disk
\be
X^2+Y^2\leq1,  \hs{3ex}\mbox{at $U=0$},
\ee
which is usually called the Poincar\'e disk.
With the disk coordinates  \(x_D,~y_D\),
we can read the relations between the points on the hyperboloid and the disk
\bea
x_D=\frac{X}{U+1}=\frac{\sinh\rho\cos\theta}{\cosh\rho+1},\nn\\
y_D=\frac{Y}{U+1}=\frac{\sinh\rho\sin\theta}{\cosh\rho+1}.\label{Diskco}
\eea
We may introduce the polar coordinates on the disk
\be
x_D=r\cos\vartheta, \hs{3ex}y_D=r\sin\vartheta.\label{Diskpolco}
\ee
Then we can solve that \(\vartheta=\theta\), and get the metric of Poincar\'{e} disk
\be
ds^2=4\frac{dr^2+r^2d\vartheta^2}{(1-r^2)^2}=4\frac{dx_D^2+dy_D^2}{(1-x_D^2-y_D^2)^2}.\label{Disk}
\ee

On the other hand, the two-dimensional de Sitter spacetime can be embedded into the same spacetime (\ref{3Dflat}) as well.
It is defined by the relation
\be
X^2+Y^2-U^2=1. \label{dSembed}
\ee
By defining a new coordinate system with the following relation
\be
U=\sinh\tau,~~~X=\cosh\tau\cos\theta,~~~Y=\cosh\tau\sin\theta,\label{dSco}
\ee
 we can get the metric of dS$_2$
\be
ds^2=-d\tau^2+\cosh^2\tau d\theta^2.\label{dS}
\ee
If we make another coordinate transformation
\be
\cosh\tau=\frac{1}{\sin\alpha},\hs{3ex}\alpha\in(0,\pi).\label{kineco}
\ee
then in terms of the coordinates $(\theta, \alpha)$ we find another metric form of the dS$_2$ spacetime
\be
ds^2=\frac{-d\alpha^2+d\theta^2}{\sin^2\alpha}. \label{kinematic}
\ee
And the point $(\theta, \alpha)$ in this coordinate will correspond to
$\displaystyle(-\cot\alpha,\frac{\cos\theta}{\sin\alpha},\frac{\sin\theta}{\sin\alpha})$ in $(U,X,Y)$ coordinate.

We can  define $H_2$ and dS$_2$ on the  Poincar\'{e} upper half plane by introducing
\be
x=\frac{X}{U-Y},\hs{3ex}y=\frac{1}{U-Y}.\label{planeco}
\ee
Then the metrics of the hyperbolic space and de Sitter spacetime are respectively
\be
ds^2=\frac{dx^2+dy^2}{y^2},\hs{3ex}\text{and}\hs{3ex}ds^2=\frac{dx^2-dy^2}{y^2}.\label{plane}
\ee

Let us define \be
z_D=x_D+\mathrm{i}y_D, \hs{3ex} z=x+\mathrm{i}y,
\ee
then the transformation between the  coordinates on the Poincar\'{e} upper half plane  and
the ones on the Poincar\'{e} disk is
\be
z=\frac{z_D+\mathrm{i}}{\mathrm{i}z_D+1},\hs{6ex}z_D=\frac{\mathrm{i}z+1}{z+\mathrm{i}}.\label{cptod}
\ee
Or, more explicitly,
\bea
  & x=\dfrac{2x_D}{x_D^2+(1-y_D)^2}, \hs{3ex} & y=\frac{1-x_D^2-y_D^2}{x_D^2+(1-y_D)^2},\nn\\
  & x_D=\dfrac{2x}{x^2+(1+y)^2}, \hs{3ex} & y_D=\frac{x^2+y^2-1}{x^2+(1+y)^2}.\label{ptod}
\eea

\subsection{Geodesics in $\mathbb{H}_2$ }

\begin{figure}
  \centering
  \includegraphics[width=0.28\linewidth]{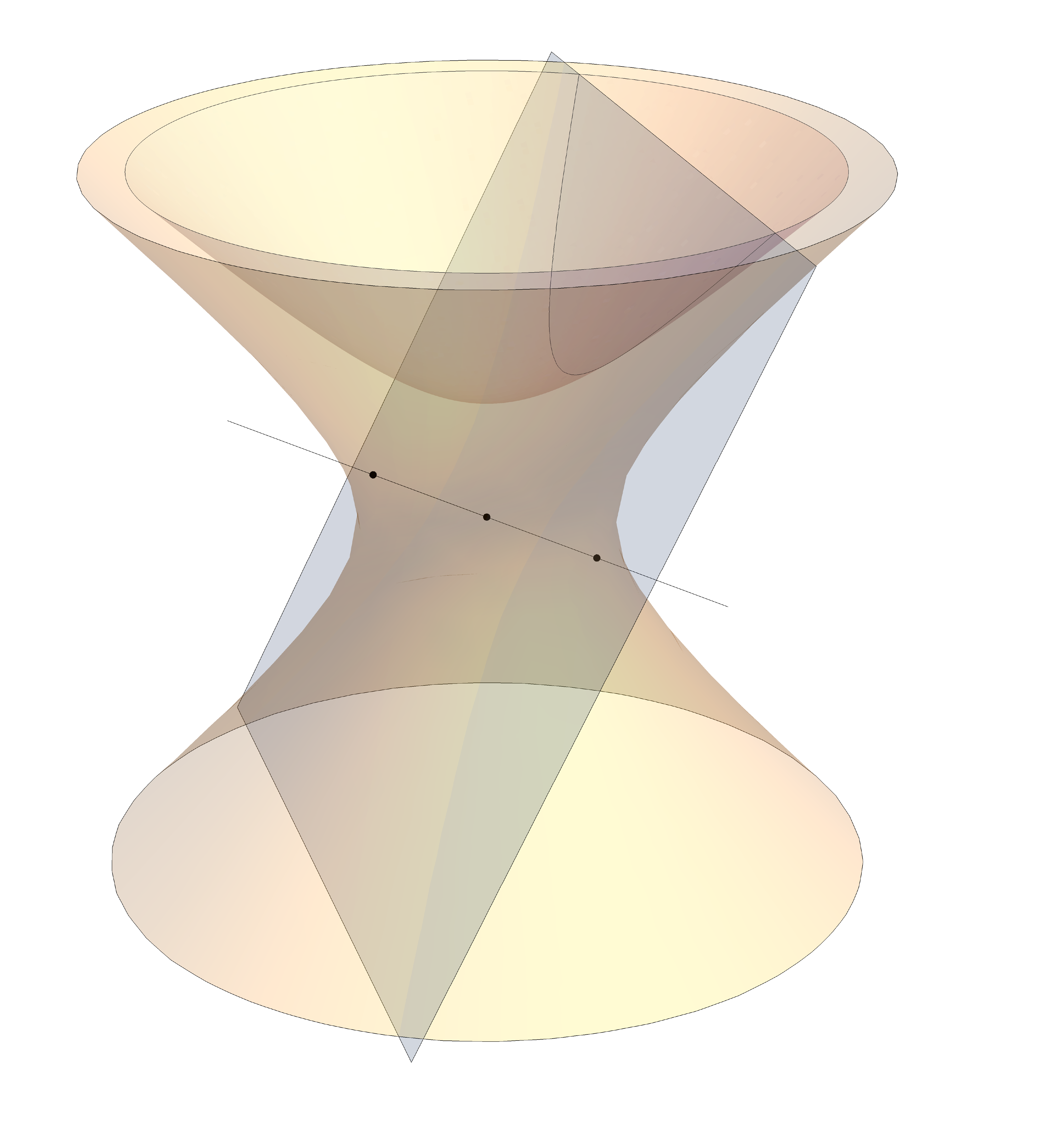}
  \includegraphics[width=0.3\linewidth]{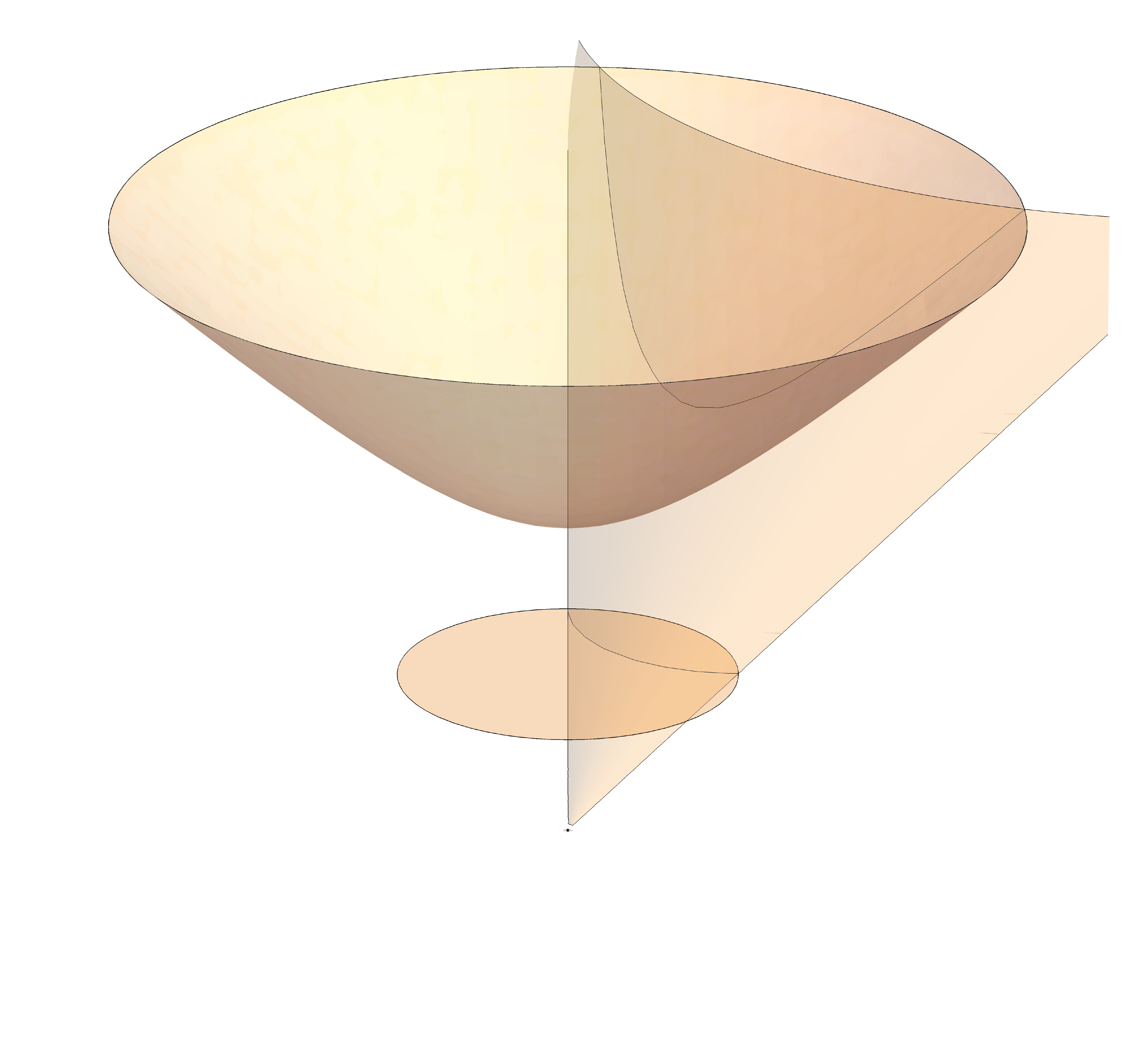}
  \includegraphics[width=0.33\linewidth]{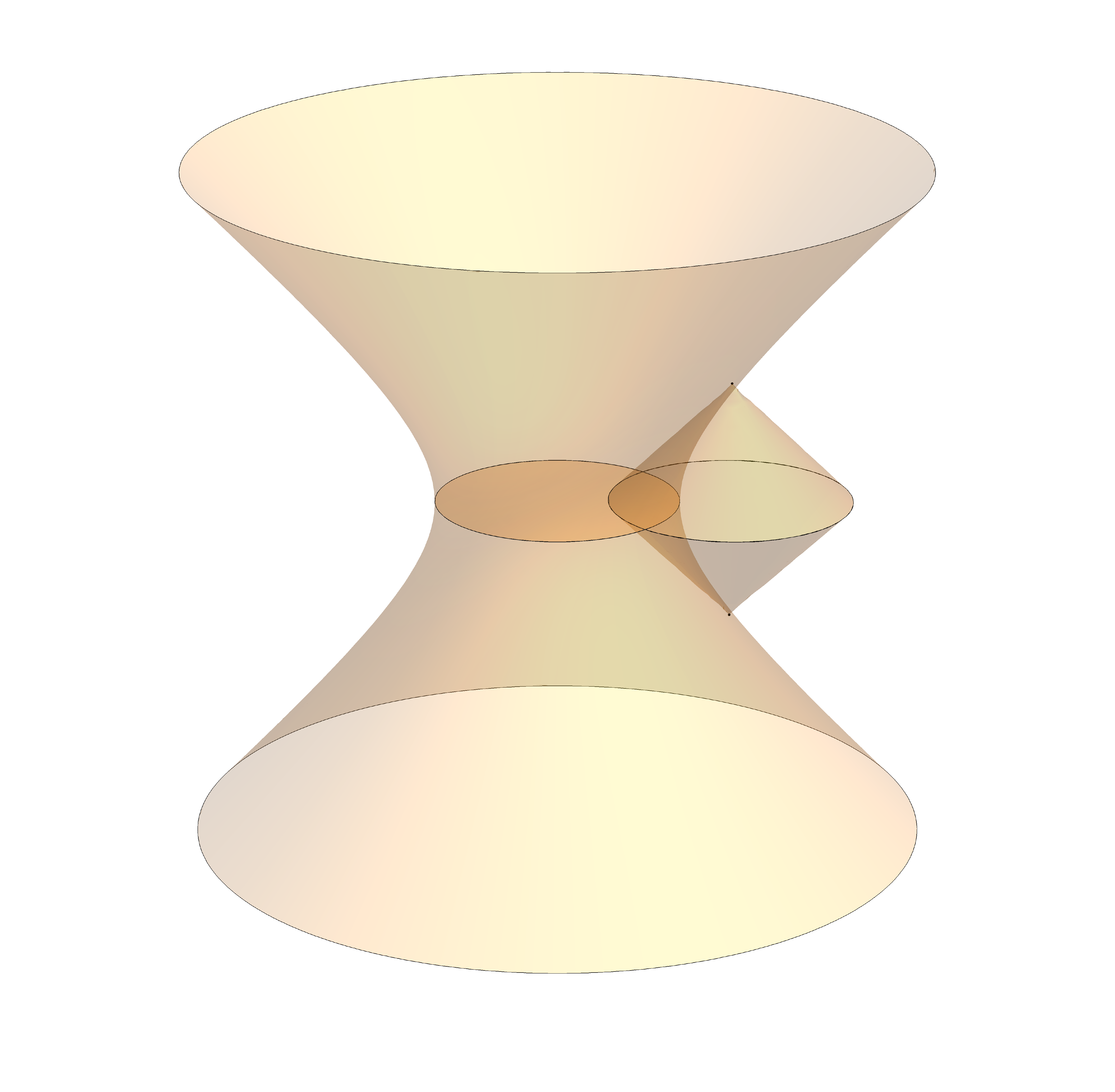}
  \caption{The upper hyperboloid is the $\mathbb{H}_2$ described by the embedding (\ref{H2co}).
  The outer one-sheeted hyperboloid is the kinematic space.
  The unit disk in the center is the Poincar\'{e} disk.
  In the left figure, the plane crossing the origin intersects $\mathbb{H}_2$ on a curve,
  which is a geodesic in $\mathbb{H}_2$.
  The line orthogonal to the plane and crossing the origin intersects the kinematic hyperboloid with two points,
  corresponding to the geodesics with different orientations.
  In the middle figure, we show the projection from the $\mathbb{H}_2$ hyperboloid to the Poincar\'{e} disk.
  The geodesic is mapped to an arc of  a circle in the disk.
  In the right figure, we show that the the future and the past domains of dependence of the disk,
  whose boundary circle intersects  the Poincar\'{e} disk with the arc,
  form a causal diamond with its tips being in the kinematic space.
  This gives another geometric construction of the kinematic space. }
  \label{hyperboloid}
\end{figure}

The geodesics in $\mathbb{H}_2$  are simple.
On $\mathbb{H}_2$, the equation of a geodesic without orientation is
\be
\tanh\rho\cos(\theta-\theta_0)=\cos\alpha_0\label{H2geo},\hs{3ex}\theta_0\in(0,2\pi),\hs{3ex}\alpha_0\in(0,\frac{\pi}{2}).
\ee
In the coordinates of $\mathbb{R}^{2,1}$, this is
\be
\cos\alpha_0U-\cos\theta_0X-\sin\theta_0Y=0. \label{3Dgeo}
\ee
This is a plane crossing the origin.
So for any geodesic on $\mathbb{H}_2$ we can find a corresponding plane crossing the origin,
and the intersection curve between this plane and hyperboloid $\mathbb{H}_2$ is just the geodesic.
The line normal to the plane and crossing the origin intersect the dS$_2$ spacetime (\ref{dSembed}) at two points,
as shown in the left of Fig. \ref{hyperboloid}.
The coordinates of these points in terms of $(U,X,Y)$  are
\be
\mp\left(\cot\alpha_0,-\frac{\cos\theta_0}{\sin\alpha_0},-\frac{\sin\theta_0}{\sin\alpha_0}\right),\label{linepoint}
\ee
In terms of $(\theta,\alpha)$ coordinate, these two points are at $(\theta_0,\alpha_0)$ and $(\pi+\theta_0,\pi-\alpha_0)$ respectively,
corresponding to the geodesics with opposite orientations.
The first point correspond to geodesic starting from $\theta_0-\alpha_0$ and ending on $\theta_0+\alpha_0$ on the boundary,
and the second point correspond to geodesic starting from $\theta_0+\alpha_0$ and ending on $\theta_0-\alpha_0$ on the boundary.

In the Poincare upper plane coordinate for $\mathbb{H}_2$,
the geodesic equation corresponding to (\ref{H2geo}) is
\be
(\cos\alpha_0-\sin\theta_0)(x^2+y^2)-2\cos\theta_0x+\cos\alpha_0+\sin\theta_0=0.\label{planegeo}
\ee
It is either a semicircle or a straight line normal to the $x$-axis
\be\left\{
\begin{array}{lc}
\mbox{Semicircle centered at $(\frac{\cos\theta_0}{\cos\alpha_0-\sin\theta_0},0)$
with radius $\left|\frac{\sin\alpha_0}{\cos\alpha_0-\sin\theta_0}\right|$,}& \cos\alpha_0\neq\sin\theta_0, \\
\mbox{Straight line normal to the $x$-axis at $x=\tan\theta_0$,}&\cos\alpha_0=\sin\theta_0
\end{array}
\right.\ee

In the Poincar\'{e} disk, as shown in the middle of Fig. \ref{hyperboloid},
the geodesic equation corresponding to (\ref{H2geo}) is
\be
\cos\alpha_0(x_D^2+y_D^2+1)-2\cos\theta_0x_D-2\sin\theta_0y_D=0,\label{Diskgeo}
\ee
The geodesic is either an arc of a circle which is orthogonal to the unit circle when $\alpha_0\neq\frac{\pi}{2}$,
or a line crossing the origin when $\alpha_0=\frac{\pi}{2}$.

The geometric meaning of $\alpha_0$ and $\theta_0$ is clear:
$\alpha_0$ is the opening angle of the arc of the unit circle intersected by the geodesic,
and $\theta_0$ is the angular coordinate of the midpoint of this arc.
In the disk, we can also denote each geodesic by the angular coordinates $(\mu,\nu)$ of its two endpoints on the unit circle,
then we may have
\be
\mu=\theta_0-\alpha_0,\hs{3ex}\nu=\theta_0+\alpha_0,\label{Diskendp}
\ee
to define the kinematic space\cite{Czech:2015qta}.
We should notice that in the kinematic space the points $(\theta_0,\alpha_0)$ and $(\theta_0+\pi,\pi-\alpha_0)$
denote the same geodesic but with different orientations.
Remarkably, the kinematic space is exactly the dS$_2$ spacetime (\ref{kinematic}) with the coordinates \((\theta,\alpha)\) and the metric (\ref{kinematic}) given above.
Therefore we can conclude that the dS$_2$ spacetime defined by (\ref{dSembed}) is exactly the kinematic space of $\mathbb{H}_2$.

In the above discussion, we have the picture that the corresponding points of a geodesic in the kinematic space
are the same as the points we get on dS$_2$ in Eq. (\ref{linepoint}) by the intersection of the  normal line to the plane (\ref{3Dgeo}).
This picture shows explicitly the relation between a hyperbolic space
and its kinematic space\footnote{This has already been pointed out in Fig. 15 in \cite{Czech:2015qta}.}.
However, in the kinematic space, the points could be timelike or spacelike separated,
depending on whether the corresponding geodesics have intersection or not\cite{Czech:2015kbp}.
It is not clear to see why there exist such a kind of relations in the above construction.

There is another geometric construction to show the causal relation of the points in the kinematic space more clearly.
As we shown above, the geodesic (\ref{planegeo}) in the Poincar\'{e} disk is actually part of a circle.
This circle is the boundary of a disk, which in general is not of unit radius.
The interesting point is that the future and the past
domains of dependence of this disk form a causal diamond with its tips being actually in the kinematic space,
as showed in the right figure of Fig. \ref{hyperboloid}. In the embedding space, the coordinates of the tips are
\be
\left(\pm\tan\alpha_0,\frac{\cos\theta_0}{\cos\alpha_0},\frac{\sin\theta_0}{\cos\alpha_0}\right),\label{conepoint}
\ee
while in the kinematic space, their corresponding coordinates \((\tilde\theta,\tilde\alpha)\) satisfy the relation
\begin{equation}
\tilde\theta=\theta_0,\hs{3ex}\tilde\alpha=\frac{\pi}{2}\pm\alpha_0.\label{conecoord}
\end{equation}
They are slightly different from the points \((\theta_0,\alpha_0)\) or \((\theta_0+\pi,\pi-\alpha_0)\)
corresponding to the geodesic by using the normal line.

However, the difference is just a constant translation.
It is the same as the kinematic space.
Therefore, we can safely take the tips of the diamond as the points corresponding to the geodesic.

This picture has the advantage to see the causal structure clearly.
For example, in the Poincar\'e disk, if two geodesics have no intersection but have the same orientation,
then the casual diamond of the outer geodesics encloses the one of the inner geodesics,
such that the corresponding point of the inner geodesic is at the casual past of the one of the outer geodesic.
This shows that the causal relation can be seen directly from the embedding picture by the relation of the corresponding light cone.
If the causal diamond of two geodesic has no intersection,
the corresponding geodesics  have no intersection as well.
And if two causal diamond have intersection, then the geodesics will also have intersection.

Moreover, in the first picture, we must decide the embedded dS$_2$ surface first, then we can get the corresponding point.
But in the second picture, we do not need to know the surface of kinematic space.
We can directly get corresponding points of all geodesics, which form the kinematic space.
And then we can get the induced metric on this surface, and this is exactly the metric of the kinematic space.

\section{Symmetries on AdS$_3$ and its quotients}

Every classical solution in the AdS$_3$ graviy is locally AdS$_3$.
They could be constructed by the quotient identification of global AdS$_3$.
For example, the BTZ geometry is a quotient of AdS$_3$ by a discrete subgroup of $PSL(2,\mathbb{R})$\cite{Banados:1992wn,Banados:1992gq}.
It is a two-boundary wormhole, or an eternal black hole\cite{Maldacena:2001kr}.

More interesting, there exist many different kinds of multi-boundary wormholes with different topology.
For the static spacetime, one may identify the geodesics in the Poincar\'e disk to construct such multi-boundary wormholes.
The detailed construction could be found in \cite{Aminneborg:1997pz,Brill:1998pr,Krasnov:2000zq,Skenderis:2009ju,Balasubramanian:2014hda,Marolf:2015vma}.
In this section, we will give a brief review of these solutions and discuss three examples carefully,
they are BTZ, three-boundary wormhole and single-boundary torus wormhole.

\subsection{Fuchsian group and its action}

In this subsection, let us focus on the symmetry transformation on the constant time slice of AdS$_3$.
For simplicity, we start from the Poincar\'e upper plane.
The symmetry group is $PSL(2,\mathbb{R})=SL(2,\mathbb{R})/\{\pm 1\}$, which could be represented by a matrix
\begin{equation}
\gamma=\left(
\begin{aligned}
a~~b\\
c~~d\\
\end{aligned}
~\right),\label{SL2Rele}
\end{equation}
with \(|\gamma|=ad-bc=1,~a,b,c,d\in\mathbb{R}\).
We require the transformation to be hyperbolic to avoid the orbifold singularities. This requirement leads to
\(|\Tr\gamma|=|a+d|>2\), which defines the Fuchsian group of the second kind.
On the half plane, we have the complex coordinate $z=x+iy$.
A point $z=x+\mathrm{i}y$ is transformed into $z^\prime=x^\prime+\mathrm{i}y^\prime=\gamma z$ under the Mobius transformation $\gamma$
\be
z^\prime=\frac{az+b}{cz+d}.
\ee
Such a transformation leads to a Riemann surface  $\Sigma = \mathbb{H}_2/ \Gamma$,
where $\Gamma$ is a discrete subgroup which is called the Fuchsian group and is generated by its fundamental element $\gamma$ as $\Gamma=\{\gamma^n|n\in\mathbb{Z}\}$.

For each transformation, we also have an one-parameter family of flow lines
\be
f(x,y)=cx^2+(d-a)x+cy^2+ey-b=0,~~~e\in\mathbb{R}.\label{flowline}
\ee
These flow lines are the integral curve of the transformation.
Every flow line is a circle which crosses the two fixed points of the transformation on the boundary $y=0$,
and $x=\frac{a-d\pm\sqrt{(a+d)^2-4}}{2c}$.
When \(e=0\), the flow line becomes a geodesic in $\mathbb{H}_2$, and we call it the geodesic flow line.
For every point $z$, we can find a $e$ such that the point locates on the corresponding flow line,
then the point $\gamma z$ locates on the same flow line as well.

Under a transformation $\gamma$, a geodesic $(x-x_0)^2+y^2=r^2$ changes to another geodesic $(x-x^\prime_0)^2+y^2=r^{\prime 2}$
with the parameters being
\bea
x_0^\prime&=&\dfrac{ac(x_0^2-r^2)+(ad+bc)x_0+bd}{(cx_0+d)^2-c^2r^2},\nn\\
r^{\prime 2}&=&\dfrac{r^2}{((cx_0+d)^2-c^2r^2)^2}.\label{transgeo}
\eea
Given two geodesics, the transformation relating them to each other is not unique.
If a geodesic is normal to every flow lines of a transformation,
then it is called a normal-geodesic of the transformation.
Given two geodesics without intersection, there exist many transformation that can transform one to another.
But there exist a unique transformation such that both geodesics are the normal-geodesics of this transformation.

\begin{figure}
  \centering
  \includegraphics[width=0.3\linewidth]{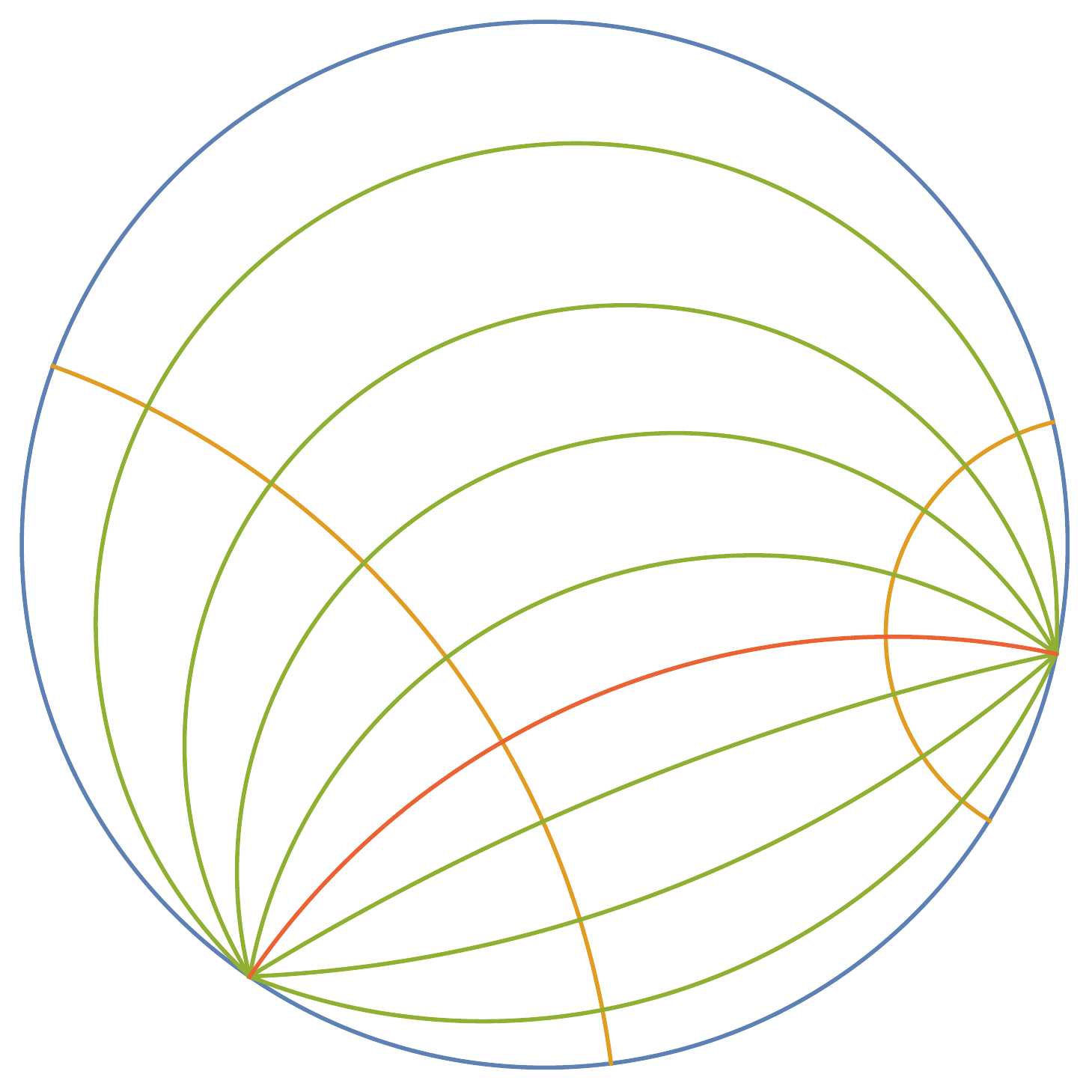}
  \caption{The unit disk inside the blue circle is the Poincar\'{e} disk.
  The two orange arcs are the normal-geodesics of a $PSL(2,R)$ transformation.
  The green and red arcs are the flow lines of this transformation, and they are normal to the two geodesics.
  The intersection point of one geodesic with a flow line is transformed into the
  intersection point of the other geodesic with the same flow line.
  Especially, the red arc is the geodesic flow line.
  The length of the arc between two intersection points of the geodesics with the geodesic flow line is the distance between  two geodesics.}
  \label{transformation}
\end{figure}

The discussion is similar in the Poincar\'{e} disk.
As shown in Fig. \ref{transformation}, among the flow lines intersecting with the geodesics, the geodesic flow line is special.
Actually, the distance between the identified points of two geodesics is the shortest along the geodesic flow line.
Such a distance is defined to be the distance of two geodesics.

\subsection{BTZ black hole as quotient}

The BTZ black hole could be taken as the quotient of the global AdS$_3$ by a discrete group of $PSL(2,\mathbb{R})$.
The action could be seen most easily in the Poincar\'{e} disk, if we are only interested in the static configuration.
In fact, if we want to get a BTZ black hole,
we should start from  a Fuchsian group defined by $\Gamma=\{\gamma^n|\gamma\in PSL(2,\mathbb{R}),n\in\mathbb{Z}\}$,
where $\gamma$ is the fundamental element such that the group is denoted as $\Gamma=\{\gamma\}$.
On the disk, we can choose a pair of non-intersecting geodesics
to be identified by this element.
Such identification can be extended to AdS$_3$ and leads to a static BTZ black hole.
The horizon length of the BTZ black hole $L_H$ can be computed directly by\cite{Maxfield:2014kra}
\be
|\Tr\gamma|=2\cosh \frac{L_H}{2}. \label{trace}
\ee
The group $\{\gamma\}$ and $\{M^{-1}\gamma M\}$ represent the same BTZ black hole.
And the flow line of the fundamental element represents the angle direction of the black hole,
while the normal-geodesics represents the radius direction.

%The parameter of wormholes can be calculated directly by just taking the trace of some proper elements in the corresponding Fuchsian group.
%Since it will be a bit more complex to get those parameters from the ,
%we can first find the transformation that maps one to the other in the normal way.

It is relatively more complex to read the horizon length from the geometric picture of identified geodesics. Let us start from a
 diagonal transformation
\be
\gamma=\left(
  \begin{aligned}
  &\lambda&0\\
  &0&\frac{1}{\lambda}
  \end{aligned}
  \right).
\ee
Its flow line is just $y=kx$, and the normal-geodesics are $\mathbf{L}:~x^2+y^2=r^2$.
The image of the normal-geodesic under the transformation is $\gamma\mathbf{L}:~x^2+y^2=\lambda^4r^2$.
Without losing generality, we can assume $\lambda>1$. In this case, the distance between two geodesics is just
\be
L_H=\int_r^{\lambda^2 r}\frac{dy}{y}=2\ln \lambda,
\ee
which is independent of the value of $r$ and match with the computation from the trace of the element (\ref{trace}).
For a general element $\gamma$, we can always find a transformation $M$ such that $\gamma^\prime=M^{-1}\gamma M$ is diagonal.
\be
\gamma=\left(
\begin{aligned}
a~~b\\
c~~d\\
\end{aligned}
~\right),~
\gamma^\prime=\left(
\begin{aligned}
\lambda~~0\\
0~~\frac{1}{\lambda}\\
\end{aligned}
~\right),~
M=\left(
\begin{aligned}
\zeta_A~~\zeta_B\\
1~~1\\
\end{aligned}
~\right).
\ee
If $\g$ is hyperbolic, then we will have $|a+d|>2$, and the above parameters are all real.
Now $\zeta_A$ and $\zeta_B$ are the coordinates of the two fixed points of $\gamma$ on the boundary for hyperbolic $\g$.
The normal-geodesic and its image are respectively
\bea
\mathbf{L}:&~\left(x-\dfrac{\zeta_Ar^2-\zeta_B}{r^2-1}\right)^2+y^2=\dfrac{(\zeta_A-\zeta_B)^2r^2}{(r^2-1)^2},\nn\\
\gamma\mathbf{L}:&~\left(x-\dfrac{\zeta_A\lambda^4r^2-\zeta_B}{\lambda^4r^2-1}\right)^2+y^2=\dfrac{(\zeta_A-\zeta_B)^2\lambda^4r^2}{(\lambda^4r^2-1)^2}.
\eea
Then we can compute the distance of any two nonintersecting geodesics described by
\be
(x-x_1)^2+y^2=r_1^2,~~~(x-x_2)^2+y^2=r^2_2,~~~r_1,r_2>0.
\ee
The four ending points of two geodesics at $y=0$ are
$u_1=x_1-r_1,~v_1=x_1+r_1,~u_2=x_2-r_2,~v_2=x_2+r_2$ respectively.
We always have $u_1<v_1,~u_2<v_2$, and without losing generality we can assume that $v_1<v_2$.
For convenience, we introduce three parameters
\be
A=(u_1-u_2)(v_1-v_2),~~~B=(u_1-v_2)(v_1-u_2),~~~C=(u_1-v_1)(u_2-v_2).
\ee
Then the horizon length of the BTZ got by identifying the two geodesics is just
\be
L_H=2\ln\left(\sqrt{\frac{|A|}{C}}+\sqrt{\frac{|B|}{C}}\right)
\ee

The above discussion can be translated into the language in the Poincar\'e disk easily.
Now in terms of the polar coordinates  $x_D=r\cos\theta,~y_D=r\sin\theta$,
the metric of the disc is of the form
\be
ds^2=4\frac{dr^2+r^2d\theta^2}{(1-r^2)^2}.
\ee
The unit circle $r=1$ is the boundary of the disk, corresponding to the boundary of $H_2$.
The points on the circle is parameterized by the angular coordinate $\theta$.
The point $(x,0)$ on the boundary of the upper half plane is mapped to the point $(1,\theta)$ with
\be
\theta=2\arctan x-\frac{\pi}{2},~~~\theta\in(-\frac{3\pi}{2},\frac{\pi}{2}).
\ee
Every geodesic in the disk can be characterized by the angular coordinates of its ending points $(\mu,\nu),~\mu<\nu$,
or equivalently in terms of the coordinates in the kinematic space $(\theta=\frac{\mu+\nu}{2},\alpha=\frac{\nu-\mu}{2})$.
For two geodesics $(\mu_1,\nu_1),~(\mu_2,\nu_2)$ in the disk, the distance between them is  just
\be\label{length of AdS}
L_H=2\ln\frac{\sqrt{|\cos(\alpha_1-\alpha_2)-\cos(\theta_1-\theta_2)|}+
\sqrt{|\cos(\alpha_1+\alpha_2)-\cos(\theta_1-\theta_2)}}{\sqrt{2\sin\alpha_1\sin\alpha_2}}
\ee

One subtle point is that  each geodesic actually corresponds to two points in the kinematic space, depending on the orientation.
The points $(\theta,\alpha)$ and $(\theta+\pi,\pi-\alpha)$ correspond to the same geodesic if we disregard its orientation.
If two points $(\theta_1,\alpha_1)$ and $(\theta_2,\alpha_2)$ are timelike separated,
then the points $(\theta_1,\alpha_1)$ and $(\theta_2+\pi,\pi-\alpha_2)$ must be spacelike separated.
The distance (\ref{length of AdS}) between two geodesics is insensitive to the relative orientation of the geodesics.
However, in order to construct the BTZ black hole by identifying the geodesics in pair, the geodesics should have the right orientations. Correspondingly the points in the kinematic space must be timelike separated.
If two points in the kinematic space are timelike separated,
then their corresponding geodesics contain each other,
have no intersection and have the same orientation.
If they are null separated, then their corresponding geodesics have one common endpoint.
And if two points are spacelike separated,
then their corresponding geodesics either have intersection or  have different orientation without intersection.

\subsection{Multi-boundary wormhole}

For the multi-boundary wormhole, the construction is similar.
Now we need more pairs of non-intersecting geodesics in the disk.
Here for simplicity, we focus on the case with  two pairs of geodesics.
With four geodesics, there exist two kinds of identification,
leading to a three-boundary wormhole and a single-boundary wormhole with the torus behind the horizon respectively.
There are two fundamental elements $\gamma_1,\gamma_2$ for the Fuchsian group $\G=\{\g_1,\g_2\}$.
The corresponding gravitational configuration is denoted as AdS$_3/\G$.

\subsubsection{Three-boundary Wormhole}

If the geodesic flow lines of the two fundamental elements do not intersect each other,
we obtain a three-boundary wormhole.
This wormhole have three asymptotic boundaries, each of which there exists a black hole.
Outside every black hole's horizon, the spacetime is described exactly by the BTZ metric.
In other words, the observer at the asymptotic infinity of each boundary sees a BTZ black hole.
Inside the horizons, the three boundaries are connected by a region with topology of a pair of pants.

The three-boundary wormhole could be characterized by  the horizon lengths $L_i$ defined on each boundary.
The horizon lengths for the first two boundaries $L_i$ are given by the $\gamma_i$: $|\Tr(\gamma_i)|=2\cosh\frac{L_i}{2}, i=1,2$,
and the horizon length for the third boundary is determined by $\gamma_3=\gamma_1\gamma_2^{-1}$.
For simplicity, we can always choose the geodesic flow lines of both transformations
$\gamma_1$ and $\gamma_2$ to be symmetric about the $x$-axis on the disc.
Moreover, we can also choose the transformation matrix of $\gamma_1$ to be diagonal.
Then we can assume  the transformation matrices are of the form
\begin{equation}\label{threehole}
  \gamma_1=\left(
  \begin{aligned}
  &\lambda~&0\\
  &0~&\frac{1}{\lambda}
  \end{aligned}\right),\hs{3ex}\gamma_2=\frac{1}{2}\left(
  \begin{aligned}
  &\left(\mu+\frac{1}{\mu}\right)+e^\alpha\left(\mu-\frac{1}{\mu}\right)~&\sqrt{e^{2\alpha}-1}\left(\mu-\frac{1}{\mu}\right)\\
  &-\sqrt{e^{2\alpha}-1}\left(\mu-\frac{1}{\mu}\right)~&\left(\mu+\frac{1}{\mu}\right)-e^\alpha\left(\mu-\frac{1}{\mu}\right)
  \end{aligned}\right),
\end{equation}
with $\lambda,\mu>1$.

\begin{figure}
  \centering
  \includegraphics[width=0.45\linewidth]{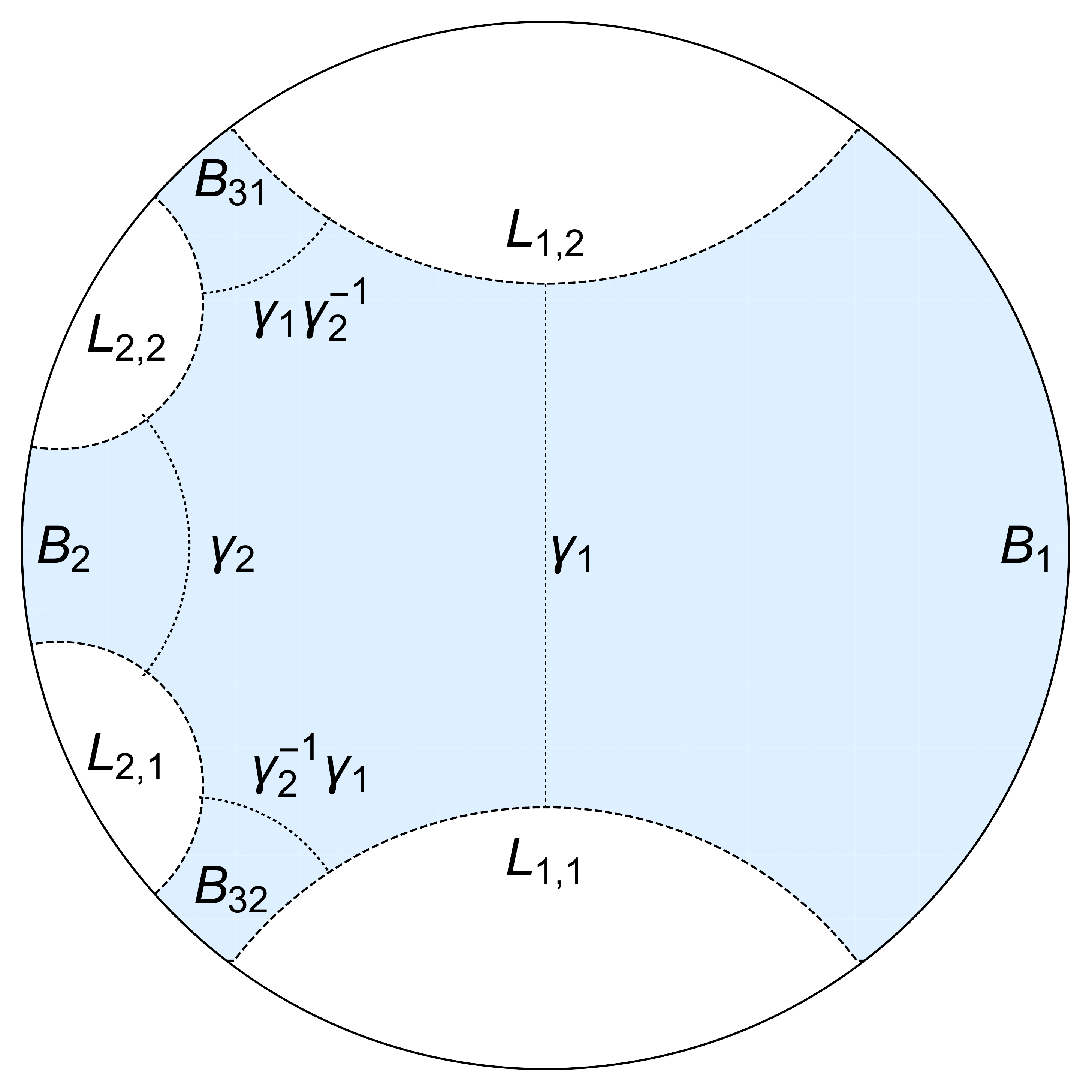}
  \includegraphics[width=0.5\linewidth]{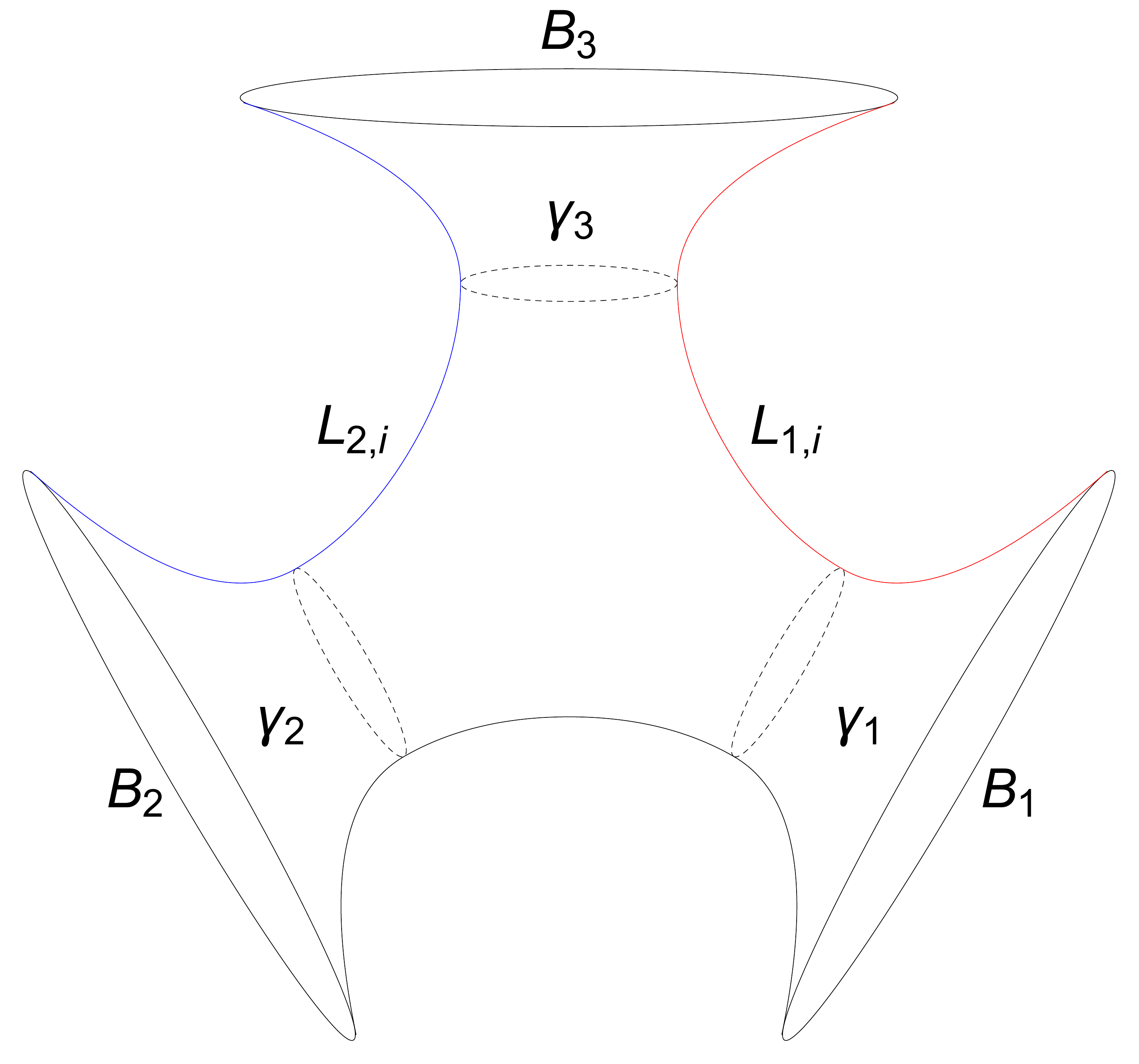}
  \caption{The three-boundary wormhole is formed by the identification of two pairs of geodesics.
  The figure on the left shows the fundamental region of three-boundary wormhole.
  The dashed lines $\mathbf{L}_{i,j}$ ending on the unit circles are normal geodesics of $\gamma_i$, and $\mathbf{L}_{i,2}=\gamma_i\mathbf{L}_{i,1}$.
  The blue region between these geodesics is a fundamental region of this wormhole.
  $B_1$ and $B_2$ denote the first two boundaries, corresponding to the transformation $\gamma_1$ and $\gamma_2$.
  $B_{31}$ and $B_{32}$ denote the two parts of the third boundary,
  which corresponds to the transformation $\gamma_3$ which is $\gamma_1\gamma_2^{-1}$ or $\gamma_2^{-1}\gamma_1$.
  And the dotted lines are the corresponding horizons.
  The figure on the right shows the topology of the $t=0$ slice of three boundary wormhole.
  The dashed lines on the right figure are the horizons of the black holes at each boundary.
  The blue and red lines are the two pair of identified normal-geodesics.}
  \label{threewormhole}
\end{figure}

The horizon lengths of the black holes on the first two boundaries are respectively
\be
L_1=2\ln\lambda, \hs{3ex}L_2=2\ln\mu. \ee
And the horizon length of the black hole on the third boundary $L_3$ is given by
\bea
\cosh\dfrac{L_3}{2}&=&\dfrac{1}{4}\left(\lambda+\dfrac{1}{\lambda}\right)\left(\mu+\dfrac{1}{\mu}\right)
-\dfrac{e^\alpha}{4}\left(\lambda-\dfrac{1}{\lambda}\right)\left(\mu-\dfrac{1}{\mu}\right)\nn\\
&=&\cosh\dfrac{L_1}{2}\cosh\dfrac{L_2}{2}-e^\alpha\sinh\dfrac{L_1}{2}\sinh\dfrac{L_2}{2},
\eea

Obviously the length $L_3$ depends on the real parameter $\alpha$, which is restricted by
\be
e^\alpha<\frac{1}{2}\left(\frac{\tanh\frac{L_1}{4}}{\tanh\frac{L_2}{4}}+\frac{\tanh\frac{L_2}{4}}{\tanh\frac{L_1}{4}}\right).
\ee

The fundamental region on the disk and the $t=0$ slice are shown on the left of Fig. \ref{threewormhole}.
The fundamental region of a three-boundary wormhole means that
every point on the disc outside this region can be mapped into it by an element of $\Gamma$.
The four geodesics are normal-geodesics of $\gamma_{1,2}$ with $\mathbf{L}_{i,2}=\gamma_i\mathbf{L}_{i,1}$.
Then by gluing each pair of normal geodesics as showed on the right of Fig. \ref{threewormhole},
we can get a surface which has the topology of a pair of pants with three boundaries.

Moreover, by acting an element $\g\in\G$, the fundamental region we chose above will be mapped to another fundamental region.
The two pairs of normal-geodesics $\mathbf{L}_{i,j}$ will be mapped to $\g\mathbf{L}_{i,j}$,
and the corresponding two fundamental elements are $\g\g_i\g^{-1}$.
Then for any fundamental region,
all of its images under the action of the elements in $\G$ can cover the whole Poincar\'{e} disc and have no intersection with each other.

\subsubsection{Torus Wormhole}

For the torus wormhole, the construction is similar as the three-boundary case,
and the only difference is that the two geodesic flow lines will intersect with each other.
This wormhole have just one asymptotic boundary with a black hole described by the BTZ metric outside the horizon.
Inside the horizon, the region has the topology of a torus with a boundary, or a pair of pants with two legs being glued together.

The torus wormhole could also be characterized by three parameters,
two of them being related to the length $L_i$ defined by the $\gamma_i$: $|\Tr(\gamma_i)|=2\cosh\frac{L_i}{2}, i=1,2$
and the other being the horizon length of the black hole defined by \(\gamma_H=\gamma_1^{-1}\gamma_2^{-1}\gamma_1\gamma_2\).

For simplity, we can choose the geodesic flow lines of both transformations $\gamma_1,\gamma_2$ to be the straight lines on the disc,
say the flow line of $\gamma_1$ being  \(x_D=0\), the one of $\gamma_2$ being \(y_D=x_D\tan\theta\).
Then we may set
\begin{equation}
  \gamma_1=\left(
  \begin{aligned}
  &\lambda&0\\
  &0&\frac{1}{\lambda}
  \end{aligned}
  \right),~\gamma_2=\frac{1}{2}\left(
  \begin{aligned}
  &\mu+\frac{1}{\mu}+\left(\mu-\frac{1}{\mu}\right)\sin\theta&\left(\mu-\frac{1}{\mu}\right)\cos\theta\\
  &\left(\mu-\frac{1}{\mu}\right)\cos\theta&\mu+\frac{1}{\mu}-\left(\mu-\frac{1}{\mu}\right)\sin\theta
  \end{aligned}
  \right).
\end{equation}
The two lengths $L_1=2\ln\lambda$ and $L_2=2\ln\mu$ characterize the torus inside the black hole.
The horizon of the black hole is determined by the element $\gamma_H$,
and the horizon length for the torus wormhole is just
\bea
  L_H&=&2\text{arccosh}\left[\frac{1}{8}\left(\lambda-\frac{1}{\lambda}\right)^2\left(\mu-\frac{1}{\mu}\right)^2\cos^2\theta-1\right],\nn\\
  &&\mbox{with}\hs{3ex}\left(\lambda-\frac{1}{\lambda}\right)\left(\mu-\frac{1}{\mu}\right)\cos\theta>4.
\eea

\begin{figure}
  \centering
  \includegraphics[width=0.45\linewidth]{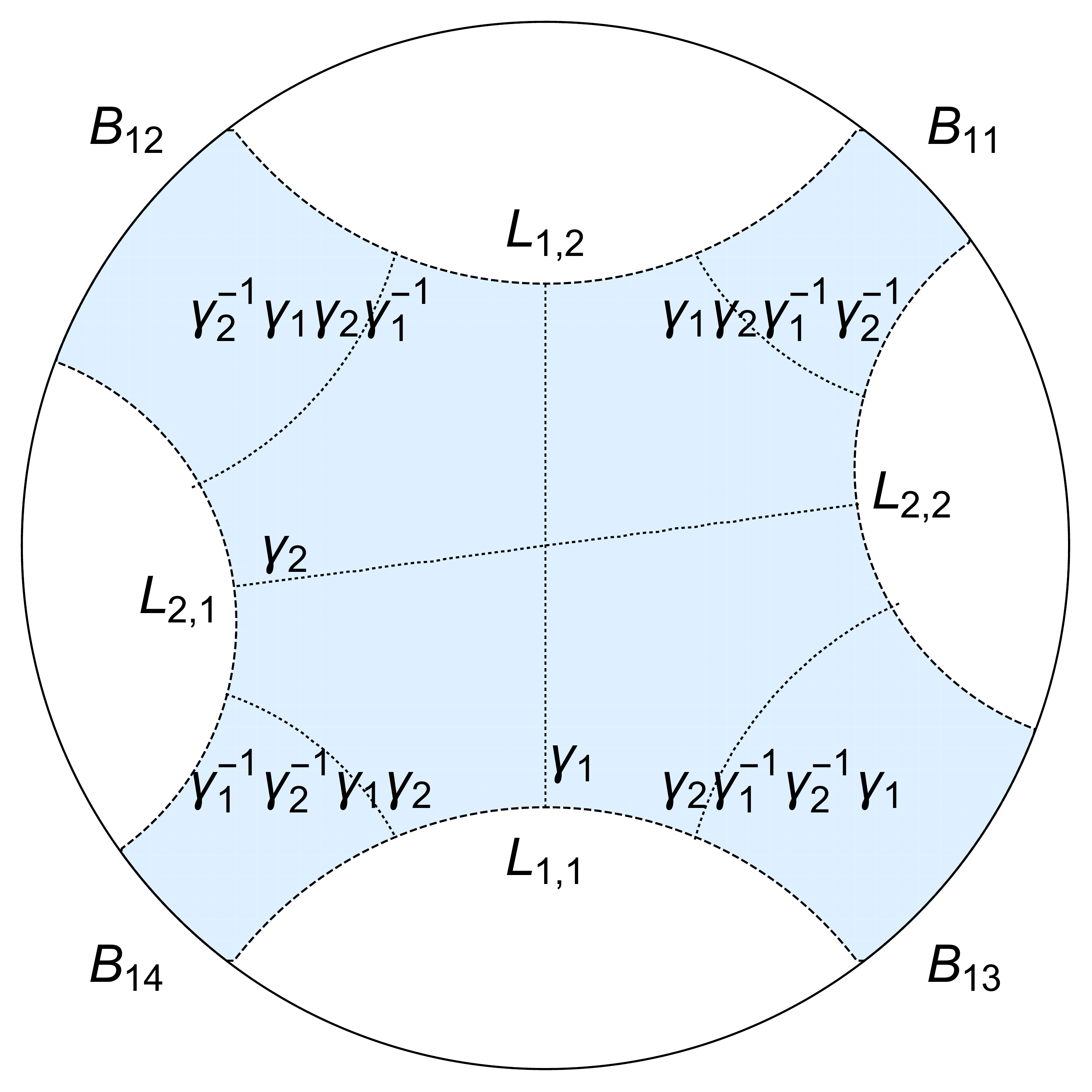}
  \includegraphics[width=0.43\linewidth]{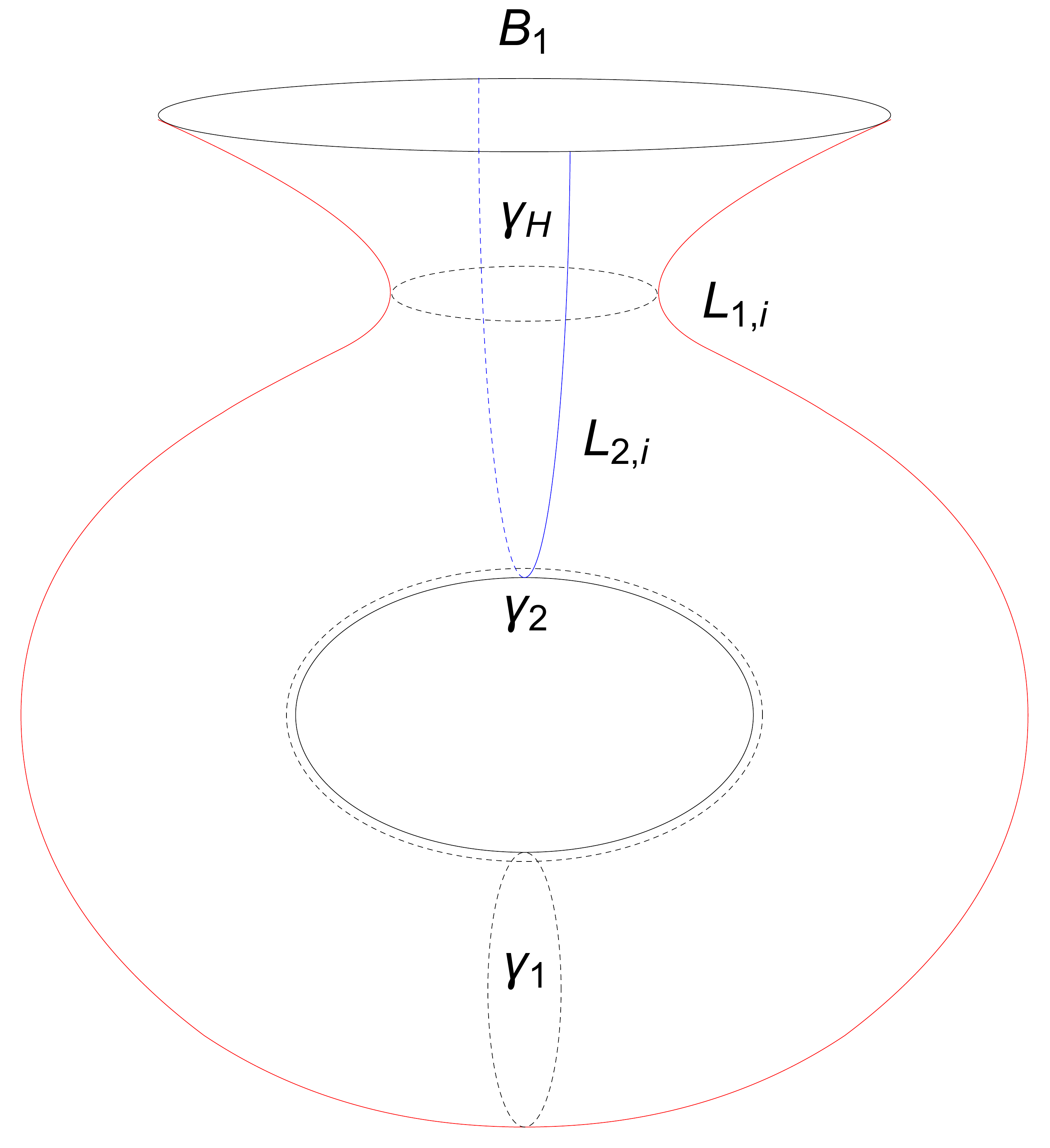}
  \caption{The torus wormhole is also given by the identification of two pairs of geodesics,
  which the flow lines of two fundamental elements $\g_1$ and $\g_2$ have a intersection.
  The figure on the left shows the fundamental region of the torus wormhole, and the marks we use here are the same as the three-boundary case.
  The identifications $\g_1$ and $\g_2$ leads to two length $L_1$ and $L_2$, which determines the shape of the torus.
  $B_{11},~B_{12},~B_{13},~B_{14}$ denote the four parts of the boundary.
  They correspond to the transformations
  $\gamma_1\gamma_2\gamma_1^{-1}\gamma_2^{-1},~\gamma_2^{-1}\gamma_1\gamma_2\gamma_1^{-1},
  ~\gamma_1^{-1}\gamma_2^{-1}\gamma_1\gamma_2,~\gamma_2\gamma_1^{-1}\gamma_2^{-1}\gamma_1$.
  These transformations are similar to each other such that they give the asymptotic boundary of the wormhole and we can mark them as $\gamma_H$.
  The figure on the right shows the topology of the $t=0$ slice of torus wormhole.
  The three dashed lines include the horizon of the black hole and two cycles with length $L_1,L_2$ that determine the region inside the horizon.
  The blue and red lines are the two pair of identified normal-geodesics.}
  \label{toruswormhole}
\end{figure}

The fundamental region on the disk and the $t=0$ slice are shown in Fig. \ref{toruswormhole}.
And the meaning of the marks and the way of construction is similar to the three-boundary case.

\section{ Kinematic space and wormhole}

In section 2, we introduced the kinematic space from a geometric point of view.
In this section, we study the properties of the kinematic space.
We discuss the geodesics in the kinematic space and show that the geodesic distance between two time-like point
is equal to the horizon length of corresponding BTZ black hole.
We also show that the normal-geodesics of a given $SL(2,\mathbb{R})$ transformation form a geodesic in the kinematic space.
Furthermore we discuss the kinematic space of the  wormholes, including the BTZ wormhole and multi-boundary wormholes.

\subsection{Geodesics in the kinematic space}

The kinematic space dS$_2$ could be described by the upper half plane $(x,y),~y\geq 0$ with the metric
\be
ds^2=\frac{-dy^2+dx^2}{y^2}.
\ee
The geodesics in it are of three types
\bea
&\text{timelike:}~~~&(x-x_0)^2-y^2=R^2,\nn\\
&\text{null:}~~~&(x-x_0)^2-y^2=0,\nn\\
&\text{spacelike:}~~~&(x-x_0)^2-y^2=-R^2.
\eea

\begin{figure}
  \centering
  \includegraphics[width=0.8\linewidth]{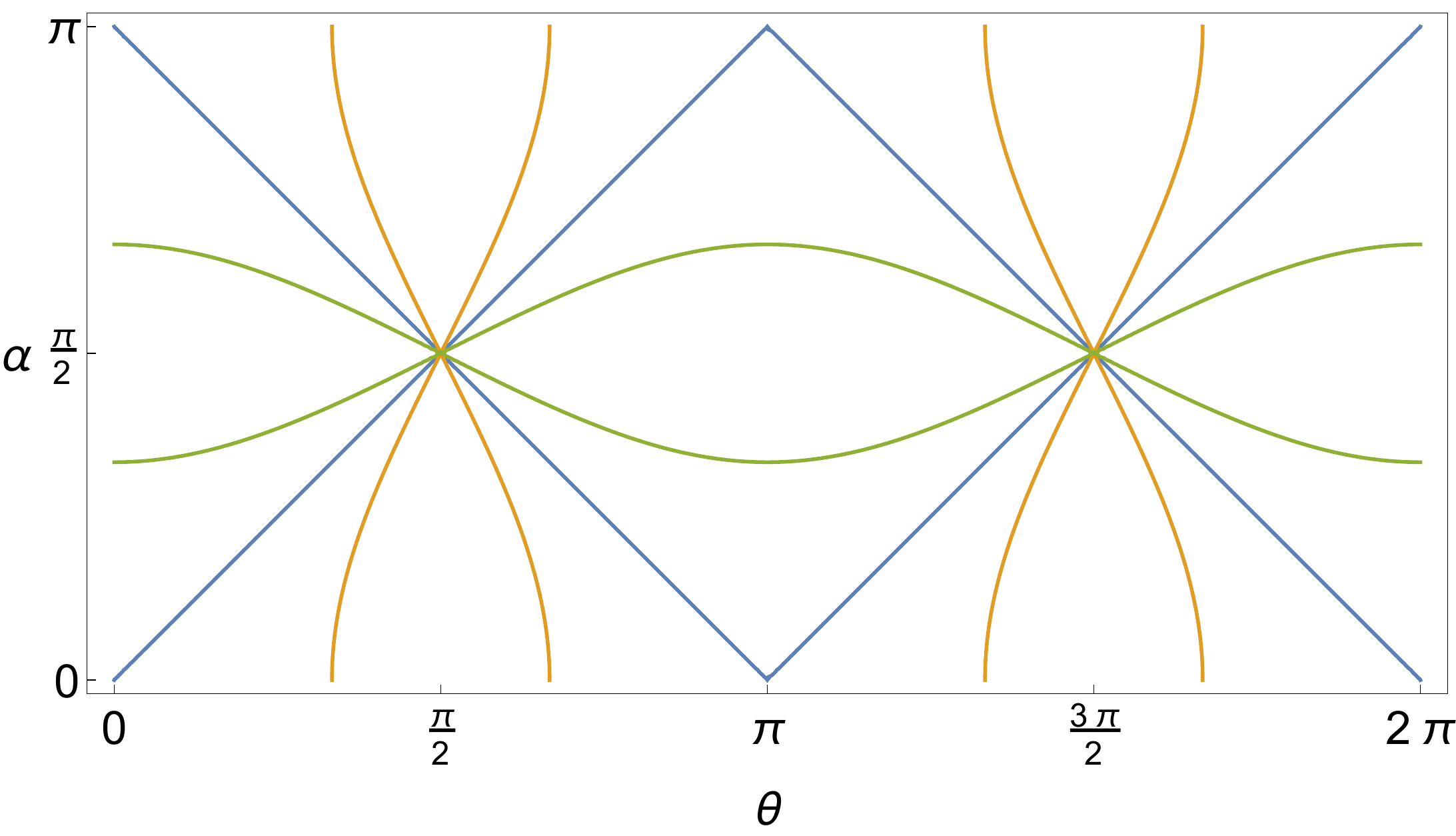}
  \caption{The geodesics in the kinematic space.
  The value of $\theta_0$ for these geodesics is taken to be $0$.
  The orange lines are the timelike geodesics with $A=\pm2$.
  The blue lines are the null geodesics with $A=\pm1$.
  The green lines are the spacelike geodesics with $A=\pm\frac{1}{2}$.}
  \label{kinematic geodesics}
\end{figure}

On the other hand, the kinematic space can be described in terms of the coordinates $(\theta,\alpha)$  with the metric (\ref{kinematic}).
Then the geodesics are described by
\be
\cos\alpha=A\cos(\theta-\theta_0),
\ee
where
\be
\left\{\begin{array}{ll}
|A|>1, \hs{3ex}&\mbox{timelike geodesic}\\
|A|=1, \hs{3ex}&\mbox{null geodesic}\\
|A|<1, \hs{3ex}&\mbox{spacelike geodesic}\end{array}
\right.
\ee
or
\be
\theta=\theta_0,
\ee
which represents a timelike geodesic.
In Fig. \ref{kinematic geodesics}, we have drawn the different geodesics in the kinematic space.

For any two timelike separated points $(\alpha_1,\theta_1),~(\alpha_2,\theta_2)$,
the geodesic connecting them has the parameters
\bea
A^2&=&\frac{\cos^2\alpha_1+\cos^2\alpha_2-2\cos\alpha_1\cos\alpha_2\cos(\theta_1-\theta_2)}{\sin^2(\theta_1-\theta_2)},\nn\\
A\cos\theta_0&=&\frac{\cos\alpha_2\sin\theta_1-\cos\alpha_1\sin\theta_2}{\sin(\theta_1-\theta_2)}.
\eea
Now the nature of the geodesic could be equivalently determined by the quantity
$\left|\frac{\alpha_2-\alpha_1}{\theta_2-\theta_1}\right|$ instead of $A^2$.
When this quantity is greater than $1$, the geodesic is timelike,
and when it is less than $1$ or equals $1$, the corresponding geodesic is spacelike or null  respectively.

The proper time between the two points along the timelike geodesic is
\bea\label{length of kinematic space}
\Delta\tau&=&\int_{\tau_1}^{\tau_2}d\tau=\int_{\alpha_1}^{\alpha_2}\frac{d\alpha}{\sin\alpha}\sqrt{1-\left(\frac{d\theta}{d\alpha}\right)^2}\nn\\
&=&\text{arctanh}\frac{\sqrt{A^2-1}\cos\alpha_1}{\sqrt{A^2-\cos^2\alpha_1}}-\text{arctanh}\frac{\sqrt{A^2-1}\cos\alpha_2}{\sqrt{A^2-\cos^2\alpha_2}}\\
&=&\text{arctanh}\frac{\sqrt{\cos^2\alpha_1+\cos^2\alpha_2-2\cos\alpha_1\cos\alpha_2\cos(\theta_1-\theta_2)-\sin^2(\theta_1-\theta_2)}\cos\alpha_1}{|\cos(\theta_1-\theta_2)\cos\alpha_1-\cos\alpha_2|}\nn\\
&&-\text{arctanh}\frac{\sqrt{\cos^2\alpha_1+\cos^2\alpha_2-2\cos\alpha_1\cos\alpha_2\cos(\theta_1-\theta_2)-\sin^2(\theta_1-\theta_2)}\cos\alpha_2}{|\cos(\theta_1-\theta_2)\cos\alpha_2-\cos\alpha_1|}
\nn\eea
This is exactly the distance between two geodesics which should be identified to obtain the BTZ black hole.
Although it seems that (\ref{length of kinematic space}) and (\ref{length of AdS}) have very different form,
 they can be proved to be equal, i.e.
 \be
\D\t\equiv L_H
\ee
Therefore we arrive the picture that the length of the horizon of the BTZ black hole
can be read from the geodesic distance of the two time-like separated points in the kinematic space,
where the two points correspond to the geodesics to be identified in the Poincar\'e disk.

\subsection{Symmetry transformation and kinematic space geodesic}

More interestingly, a given isometric transformation defines a geodesic in the kinematic space.
Let us consider a hyperbolic transformation
\begin{equation}
\gamma=\left(
\begin{aligned}
a~~b\\
c~~d\\
\end{aligned}
~\right),
\end{equation}
whose normal-geodesics in the Poincar\'e upper half plane can be parameterized by $r_0$ and are given by
\be
(x-x_P)^2+y^2=r_P^2,
\ee
with
\be
x_P=\frac{\z_Ar_0^2-\z_B}{r_0^2-1},\hs{3ex}r_P=\left|\frac{(\z_A-\z_B)r_0}{(r_0^2-1)}\right|,
\ee
where $\z_A,\z_B$ are elements in the matrix $M$.
In the disk, the normal-geodesics are given by
\be
 x^2-2x_Dx+y^2-2y_Dy+1=0,
\ee
 where
\bea
x_D=\frac{2x_P}{x_P^2-r_P^2+1}=\frac{2(\z_Ar_0^2-\z_B)}{(\z_A^2+1)r_0^2-(\z_B^2+1)},\nn\\
y_D=\frac{x_P^2-r_P^2-1}{x_P^2-r_P^2+1}=\frac{(\z_A^2+1)r_0^2-(\z_B^2-1)}{(\z_A^2+1)r_0^2-(\z_B^2+1)}.
\eea

The points in the kinematic space corresponding to the above one-parameter geodesics have $0<\alpha<\frac{\pi}{2}$, so that $\cos\alpha>0$.
The coordinates of these points are determined by the equations
\be
\tan\theta=\frac{y_D}{x_D},~~~\cos\alpha=\frac{1}{\sqrt{x_D^2+y_D^2}}
\ee
Then we find a curve in the kinematic space, which is determined by the relation
\bea
\cos\alpha=\pm\frac{\sqrt{a^2+b^2+c^2+d^2-2}}{|c-b|}\cos(\theta-\theta_0),~~~~\tan\theta_0=\frac{b+c}{d-a}
\label{trg}
\eea
This  is a timelike geodesic in the kinematic space.
On the contrary, if we require that this curve is  timelike, we should have
\be
\frac{\sqrt{a^2+b^2+c^2+d^2-2}}{|c-b|}>1
\ee
which is equivalent to
\be
|\Tr\gamma|=|a+d|>2.
\ee
In other words, the element $\gamma$ should be hyperbolic.

For hyperbolic and elliptic transformation $\g$, the normal-geodesics are
\be
(x-x_P)^2+y^2=x_P^2-\frac{a-d}{c}x_P-\frac{1-ad}{c^2}
\ee
Then we have
\bea
x_D=\frac{2c^2x_P}{c(a-d)x+ad-1+c^2},\nn\\
y_D=\frac{c(a-d)x+ad-1-c^2}{c(a-d)x+ad-1+c^2}.
\eea
In the kinematic space, the corresponding points form a geodesic, still described by Eq. (\ref{trg}).
However, the geodesic is no longer timelike.
Actually, for an elliptic  transformation the geodesic is spaclike, while for a parabolic transformation the geodesic is null.

\subsection{ BTZ and kinematic space}

We have showed that the horizon length $L_H$ in the BTZ spacetime equals  the geodesic distance $\Delta\tau$ in the kinematic space.
In the kinematic space,  for any pair of time-like separated points, it corresponds to a BTZ spacetime.
On the other hand, for a fixed BTZ spacetime obtained by the identification $\{\g\}$ on a pair of geodesics in the Poincar\'e disk,
it would be interesting to discuss its kinematic space.
The kinematic space for the BTZ spacetime is still defined by the geodesics in the BTZ spacetime.
We may start from the geodesics in the Poincar\'e disk, and take into account of  the identification $\{\g\}$ on all the geodesics.

\begin{figure}
  \centering
  \includegraphics[width=0.3\linewidth]{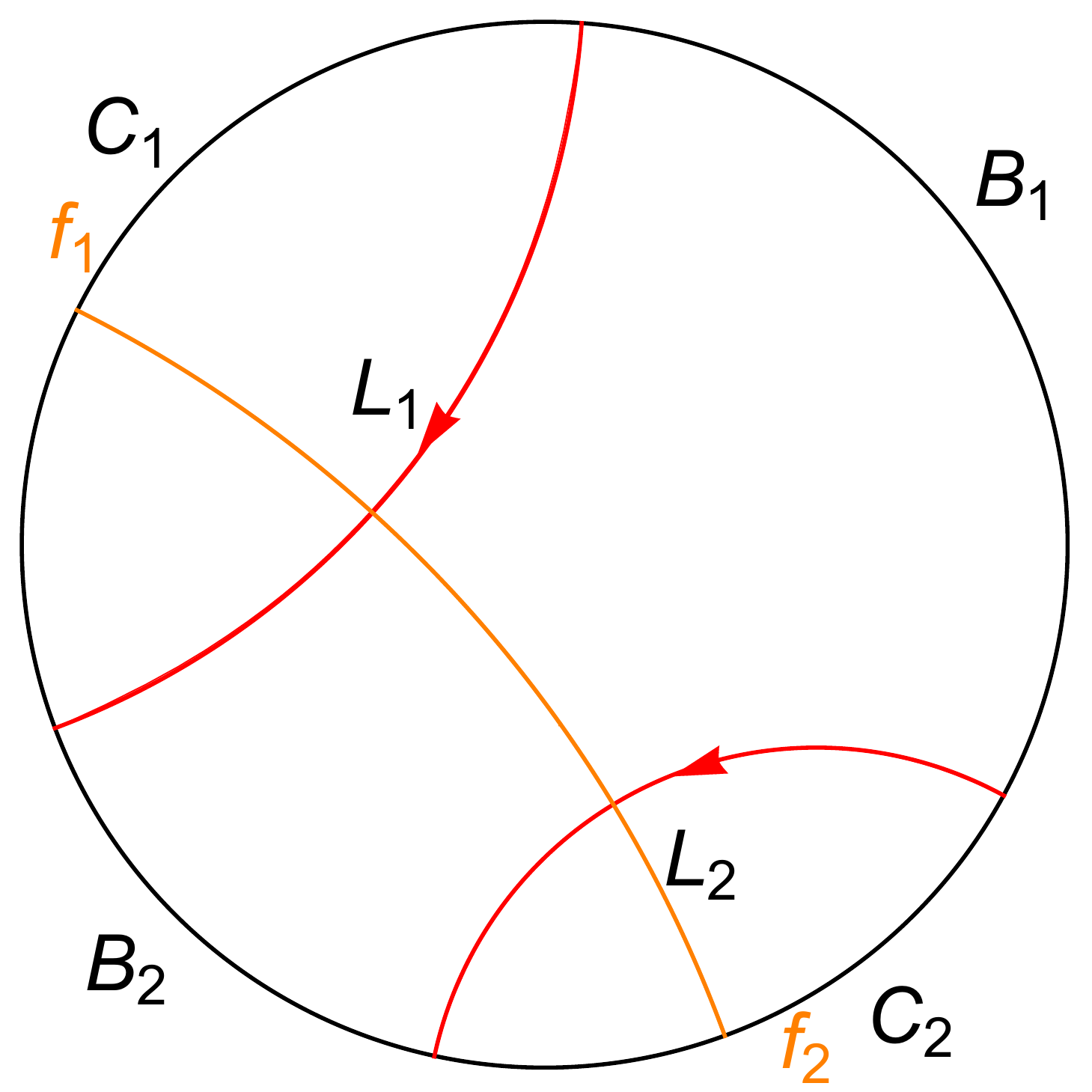}
  \includegraphics[width=0.55\linewidth]{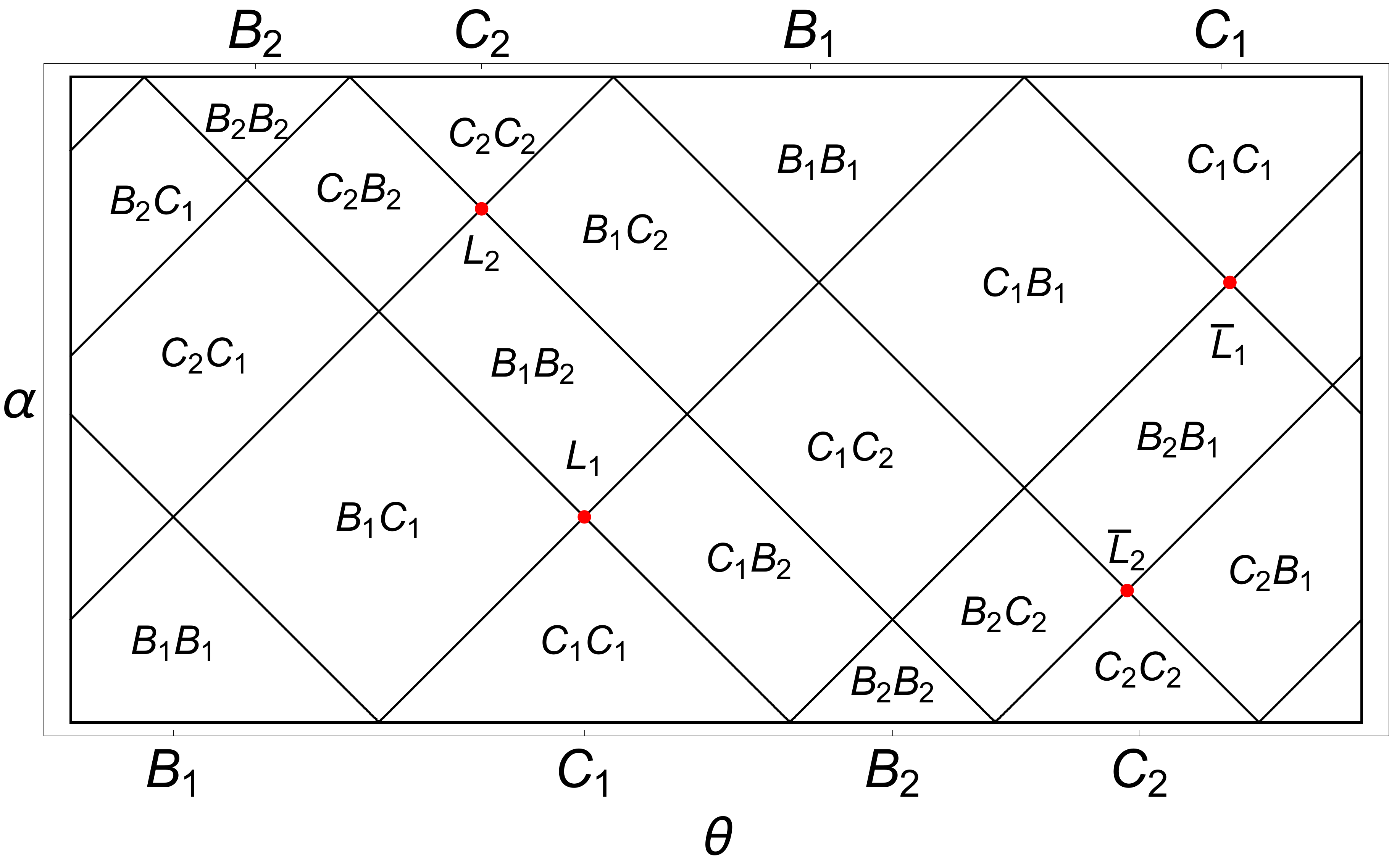}
  \caption{On the left, we draw the four regions on the boundary of Poincar\'{e} disk
  which are divided by the two oriented red geodesics identified with each other.
  The orange line is the horizon,
  and its two endpoints on the boundary is the fixed points of the corresponding transformation.
  On the right, we separate the whole kinematic space into 20 regions and marked them by the starting point and ending point of the corresponding geodesics.
  }
  \label{BTZcor}
\end{figure}

As showed in the left figure of Fig. \ref{BTZcor},
the BTZ spacetime is obtained by identifying a pair of non-intersected geodesics $\mathbf{L}_1,~\mathbf{L}_2=\gamma \mathbf{L}_1$.
Between these two geodesics, there is a fundamental region.
On the boundary, the two geodesics divide the boundary of the disk into four parts.
We mark them as $B_1,B_2,C_1,C_2$, where $B_i$'s are the boundaries of the fundamental region,
corresponding to the two boundaries of the BTZ wormhole,
and $C_i$'s are the remaining parts on the circle.
Then we can label any geodesic in H$_2$ by the regions where its two endpoints locate.
For example, a geodesic with one endpoint in $B_1$ and the other in $C_2$ is labelled by $B_1C_2$ or $C_2B_1$.
Note that the order in the label represents the orientation:
$B_1C_2$ means the geodesics have a starting point in $B_1$ and an ending point in $C_2$.
Here we should notice that a geodesic with the parameter $(\a,\theta)$ has
the starting point at $\mu=\theta-\a$ and the ending point at $\nu=\theta+\a$.

As the BTZ spacetime is a quotient of AdS$_3$ under the action of a Fuchsian group $\G$,
its kinematic space cannot be the same as the one of AdS$_3$.
If two points in the kinematic space of AdS$_3$ can be transformed to each other under the action of an element in the $\G$,
they represent the same geodesic in the BTZ spacetime.
We would like to find the fundamental region in the kinematic space of AdS$_3$, corresponding to the BTZ black hole.
Here "fundamental" means that each geodesic with orientation in the BTZ spacetime
has and only has one corresponding point in that region.
Since there are many different points representing the same geodesic up to identification,
there is ambiguity in choosing the fundamental region.
Here, we give a universal rule based on the two identified geodesics  defining the fundamental region in the Poincar\'{e} disk.

As shown in the right figure of Fig. \ref{BTZcor},
the kinematic space can be separated into 20 regions by the geodesics with different ending points and orientations.
Note that if we reverse the orientations of the $\mathbf{L}_1$ and $\mathbf{L}_2$ simultaneously,
the identification of them leads to the same BTZ black hole.
We label the points corresponding to the geodesics with opposite orientation to
$\mathbf{L}_1$ and $\mathbf{L}_2$ as $\bar{\mathbf{L}}_1$ and $\bar{\mathbf{L}}_2$.

There are two fixed points under the action $\G=\{\g\}$, as shown in the left figure of Fig. \ref{BTZcor}
which are the intersection point between the orange geodesic and the boundary.
They lie on the boundaries $C_1$ and $C_2$, labelled by $f_1$ and $f_2$.
They divides the boundaries $C_1$ and $C_2$ into two parts respectively.
The fundamental region for the BTZ spacetime in the Poincar\'e disk is the region
between two geodesics $\mathbf{L}_1$ and $\mathbf{L}_2$ with two boundaries $B_1$ and $B_2$.
Under the action of $\g$, the fundamental region is transformed to the region with the boundaries in $C_1$ next to $B_1$ and $B_2$.
Similarly the action of $\g^{-1}$ transforms the fundamental region to the region with the boundaries in $C_2$ next to $B_1$ and $B_2$.
Furthermore all the regions under  action of $\g^n, n\in \mathbb{Z}$ on the fundamental region cover the whole Poincar\'e disk.

On the other hand, each geodesics in the Poincar\'e disk can be related to the one in the BTZ spacetime by the action of $\G$.
If the ending point of the geodesic in the disc is in $C_i$, it can always be mapped to the point in $B_i$.
However, the resulting geodesic in the BTZ spacetime may wind around the horizon.
In order to classify the geodesics in the BTZ spacetime, we need to consider the action of the Fuchsian group more carefully.

\begin{figure}
  \centering
  \includegraphics[width=0.9\linewidth]{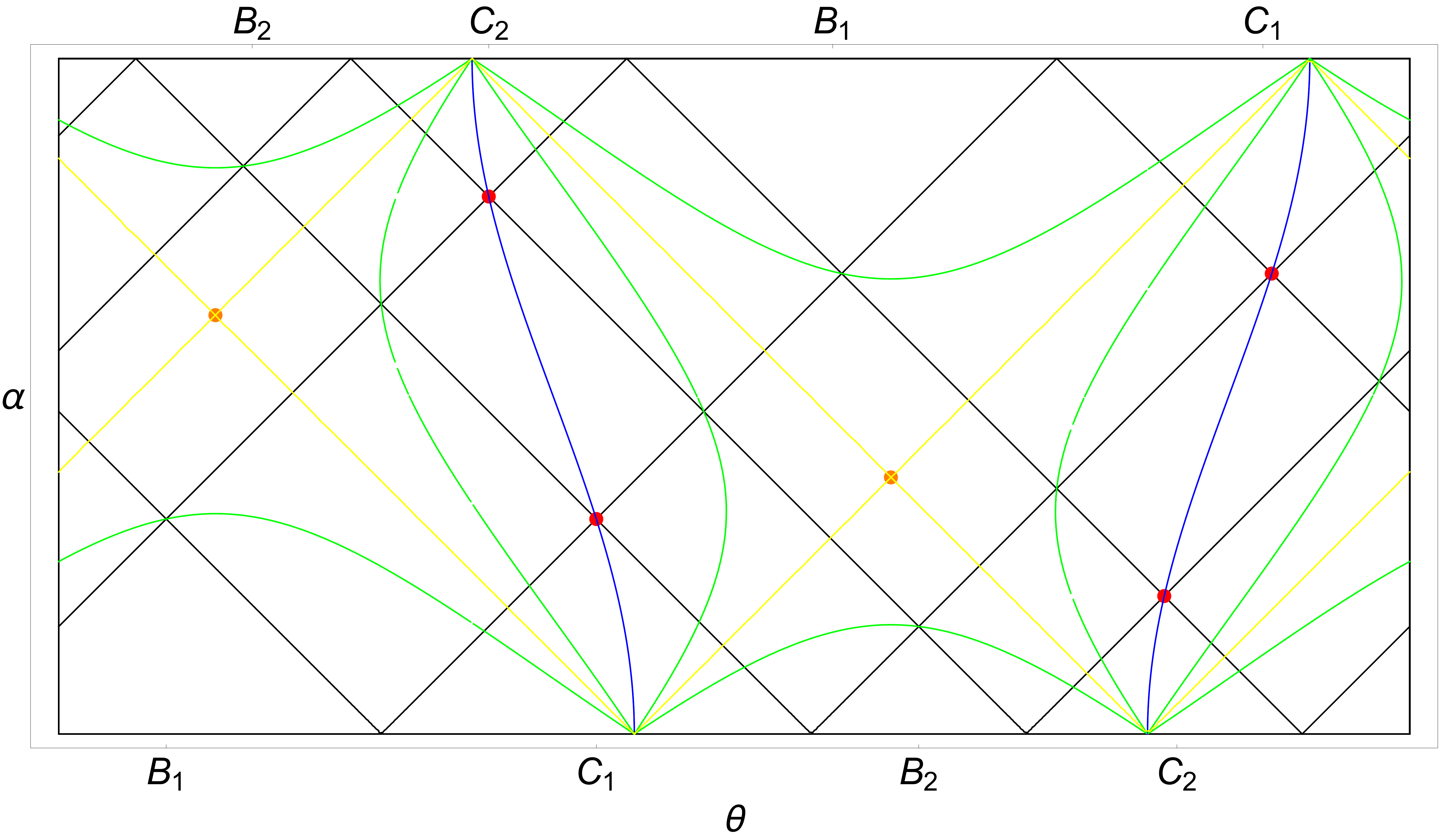}
  \caption{The blue lines are the timelike geodesic corresponding to $\G$,
  and the corresponding geodesics in $\mathbb{H}_2$ are the normal-geodesics of $\G$.
  The points on yellow lines represent the geodesics with one endpoint being the fixed point of $\G$,
  and having infinite windings around the horizon.
  The orange intersection points of two yellow lines correspond to the geodesic covering the horizon of the BTZ black hole.
  The points on the green lines represent the geodesics with two endpoints having the same angular coordinate and winding the horizon once.
  }
  \label{BTZregion}
\end{figure}

To discuss the action of the Fuchsian group on the geodesics in the disk,
we start from the geodesics with at least one ending point being the fixed point and study the action of $\G$ on them.
Actually, as any point on the boundary can be mapped to the fixed point by the continual action of the fundamental element $\g$,
all the geodesics on the Poincar\'e disk can be related to the geodesics ending at the fixed point.
In other words, starting from the geodesics ending at the fixed point and consider its images under the action of the element $\g^n, n\in Z$,
these images constitutes a line in the kinematic space, starting and ending at two fixed points.
Moreover, each fixed point with the angular coordinate $\theta$ actually corresponds to two points
with the coordinates $(\theta,0)$ and $(\theta+\pi,\pi)$ on the boundary of the kinematic space.
Therefore, as shown in Fig. \ref{BTZregion}, there are various lines, connecting two of the fixed points.
Among all the lines, there are a few special ones, which are drawn in colored lines in  Fig. \ref{BTZregion}.

The blue lines represent all the normal-geodesics of $\gamma$,
and the red points on it represent the geodesics $\mathbf{L}_1$, $\mathbf{L}_2$, $\bar{\mathbf{L}}_1$ and $\bar{\mathbf{L}}_2$ respectively.
These geodesics represent the radial direction, all the points on the same geodesic having the same angular coordinate.
Moreover, the blue lines themselves are also geodesic in the kinematic space.

The points on yellow lines represent all the geodesics with one endpoint being the fixed point and having infinite windings around the horizon.
The two intersection points between the yellow lines represent the geodesic connecting the two fixed points on the boundaries of $\mathbb{H}_2$.
The geodesic actually covers the horizon of the BTZ wormhole.
If the point in the kinematic space is timelike separated from the intersection points,
then the corresponding geodesic does not intersect with the horizon and its  endpoints are on the same boundary.
And if the point is spacelikely separated from the intersection point,
then the corresponding geodesic does intersect with the horizon and so its endpoints are on different boundaries.
Thus, the yellow lines separate all geodesics into the ones with two endpoints on the same boundary and those on different boundaries.

The points on the green lines correspond to the geodesics for which one of its endpoint
can be mapped into the other by the fundamental transformation $\g$.
Or in other words, such geodesics wind around the horizon once.
Therefore the green lines separate all geodesics into the ones with or without the winding  around the horizon.
The points in the regions between the two timelike green lines containing the blue lines
correspond to the geodesics without winding and with the endings on different boundaries.
The points in the regions between the spacelike green lines and the boundary of kinematic space
correspond to the geodesics without winding and with the endings  on the same boundary.
The points in the regions between all the green lines  containing the yellow lines correspond to the geodesics with windings on the horizon,
and the yellow lines divide them into the ones ending on different boundaries or on the same boundary.
textcolor{red}{Another important feature is that the causal relation of two points will be invariant under the action of any transformation.}

\begin{figure}
  \centering
  \includegraphics[width=0.45\linewidth]{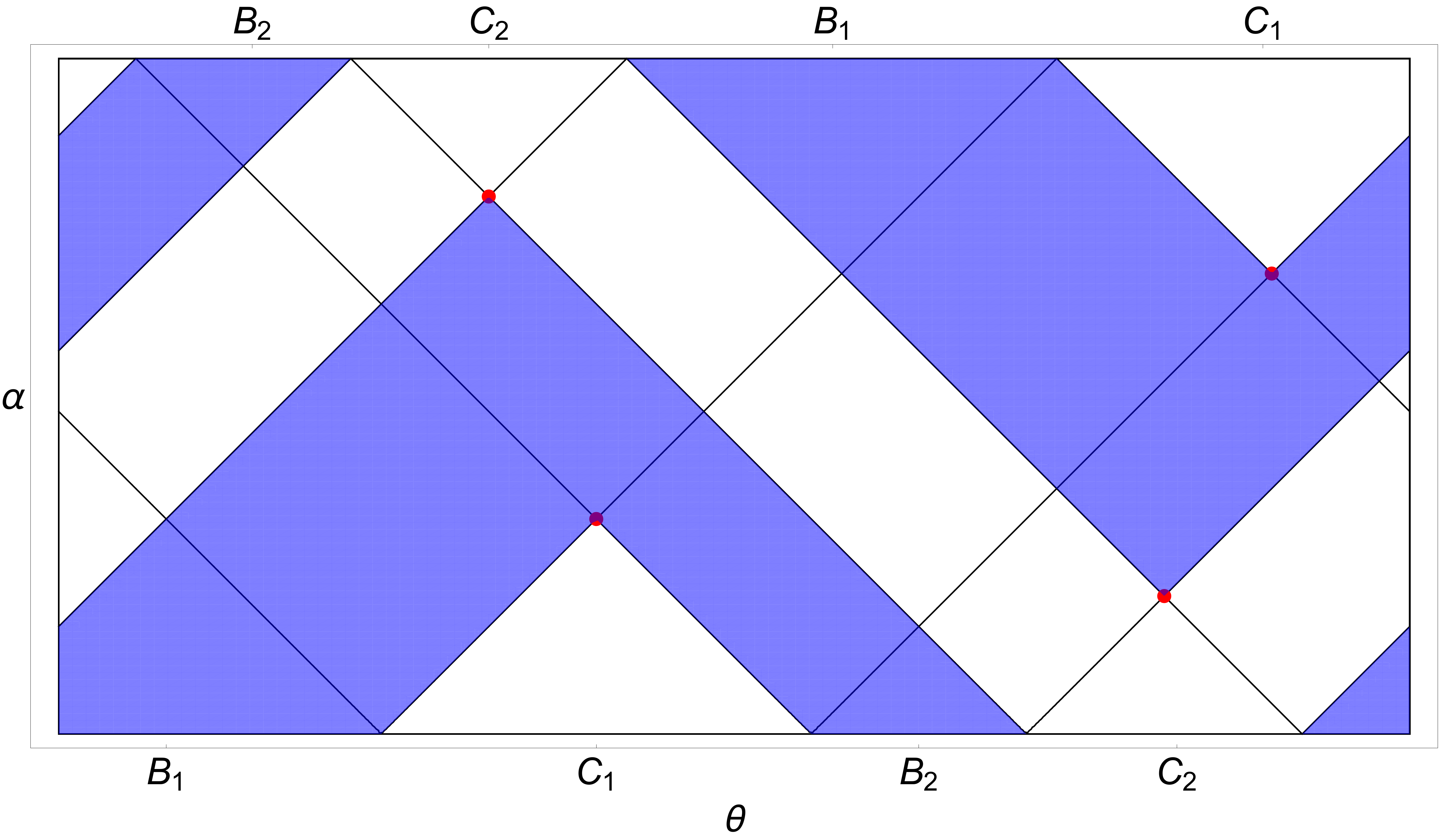}
  \includegraphics[width=0.45\linewidth]{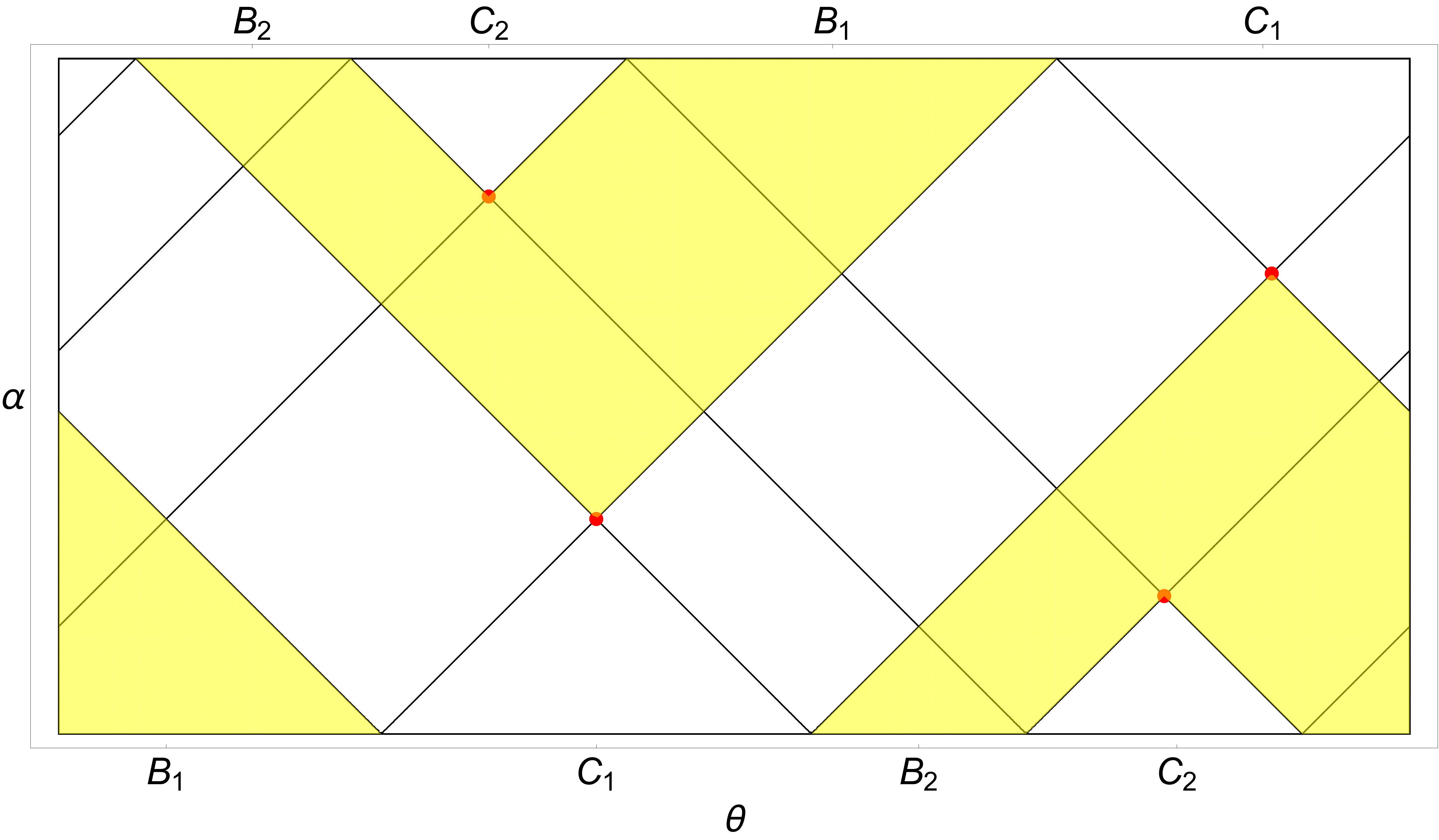}
  \caption{Two different choices for the fundamental regions for a BTZ wormhole.
  On the left we include the regions $B_iC_1$ and $C_1B_i$,
  and on the right we include the regions $B_iC_2$ and $C_2B_i$ instead.
 Both of them  includes the $B_iB_j$ regions.
  We define the left figure corresponding to the identification $\mathbf{L}_2=\g\mathbf{L}_1$,
  and the right figure corresponding to the identification $\bar{\mathbf{L}}_1=\g^{-1}\bar{\mathbf{L}}_2$.
  }
  \label{BTZfund}
\end{figure}

Now we can determine  the fundamental region of the  BTZ wormhole in the kinematic space.
The points in the regions $B_1B_1$ and $B_2B_2$ represent the geodesics ending on the same boundary,
and the ones in $B_1B_2$ represent the geodesics ending on different boundaries,
and all of them correspond to the geodesics without winding.
For a point in the region $C_1C_1$, $C_2C_2$ or $C_1C_2$,
we can always find an element in $\G$ which transforms at least one endpoint of the corresponding geodesic into $B_i$.
So the regions $C_iC_j$ will not be included into the fundamental region.
Then the main question is focused on the regions  $B_iC_j$ and $C_jB_i$.
Since the geodesics corresponding to the points in these two  regions differ  only  on orientation,
we discuss only $B_iC_j$ bellow.
For every geodesic in $B_iC_2$ we can always find an element in $\G$ which
transforms the endpoint in $B_i$ to $C_1$ and the other endpoint in $C_2$ to $B_i$,
we just need to choose the regions $C_1B_i$ and $B_iC_1$, or the regions $C_2B_i$ and $B_iC_2$, to be part of the fundamental region.
For the former choice,  the corresponding fundamental region is drawn in blue in the left figure  of Fig. \ref{BTZfund}.
These is similar as the choice in Fig.17 of \cite{Czech:2015kbp}, but with a little difference because they ignore the orientation.
For the latter choice, the fundamental region is drawn in yellow in the right figure of Fig. \ref{BTZfund}.

Here we make a rule for the choice, which will be used in the discussion of multi-boundary wormhole.
If the Fuchsian group we choose is generated by $\g$, $G=\{ \g \}$, with $\mathbf{L}_2=\g\mathbf{L}_1$,
then we choose the fundamental region to include $C_1B_i$ and $B_iC_1$.
But if the Fuchsian group we choose is generated by $\g^{-1}$, $G=\{ \g^{-1} \}$,
with $\bar{\mathbf{L}}_1=\g^{-1}\bar{\mathbf{L}}_2$,
then we choose the fundamental region to include $C_2B_i$ and $B_iC_2$.
Actually, both choices correspond to the same wormhole with the same identification,
we make this rule just for  self-consistent discussion.
It does not make any difference if we choose an opposite rule.
If we glue the points on the boundaries that are identified and has the same orientation,
then the topology of the fundamental region is just two disconnected cylinders.

\subsection{Multi-boundary wormhole in kinematic space}

\begin{figure}
  \centering
  \includegraphics[width=0.3\linewidth]{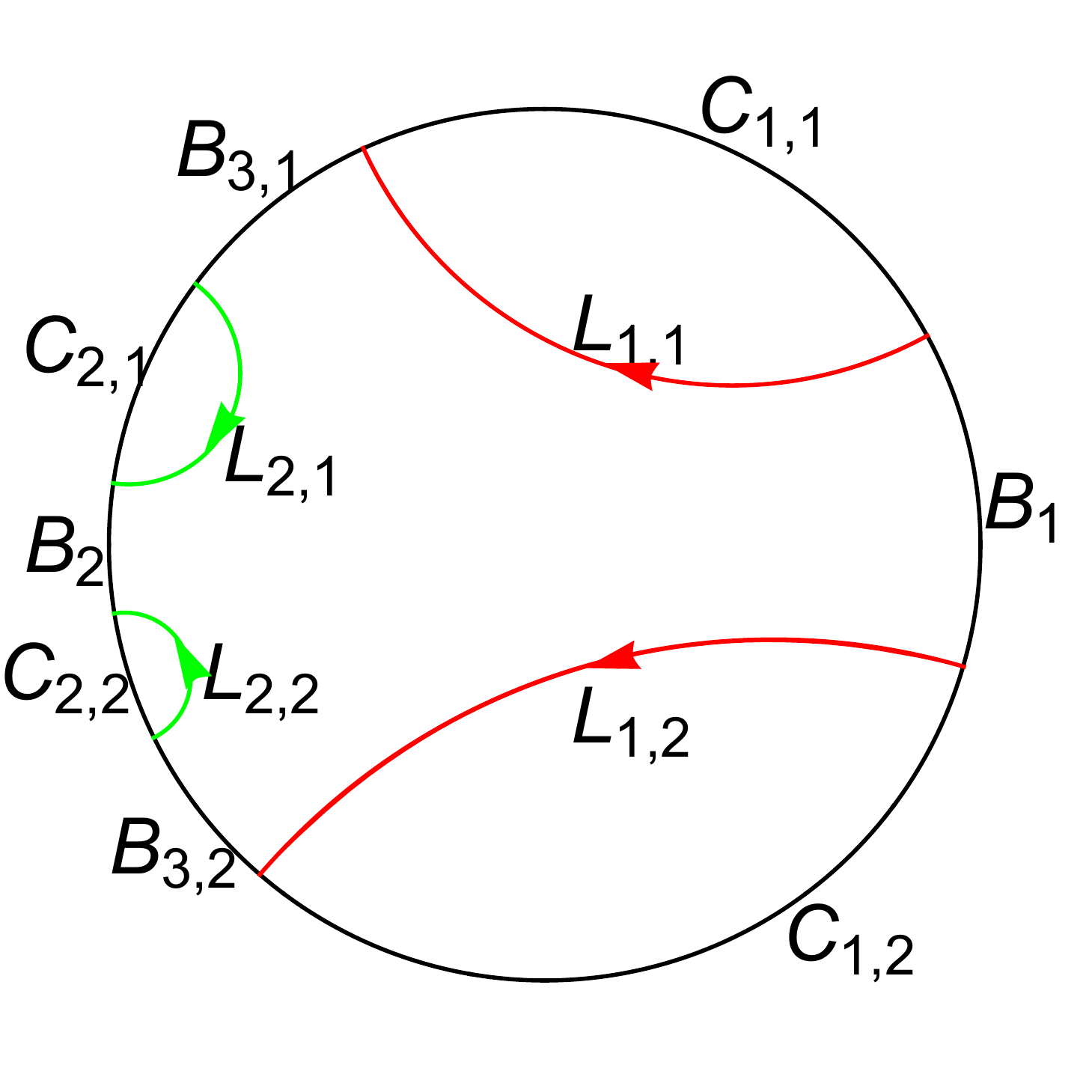}
  \includegraphics[width=0.55\linewidth]{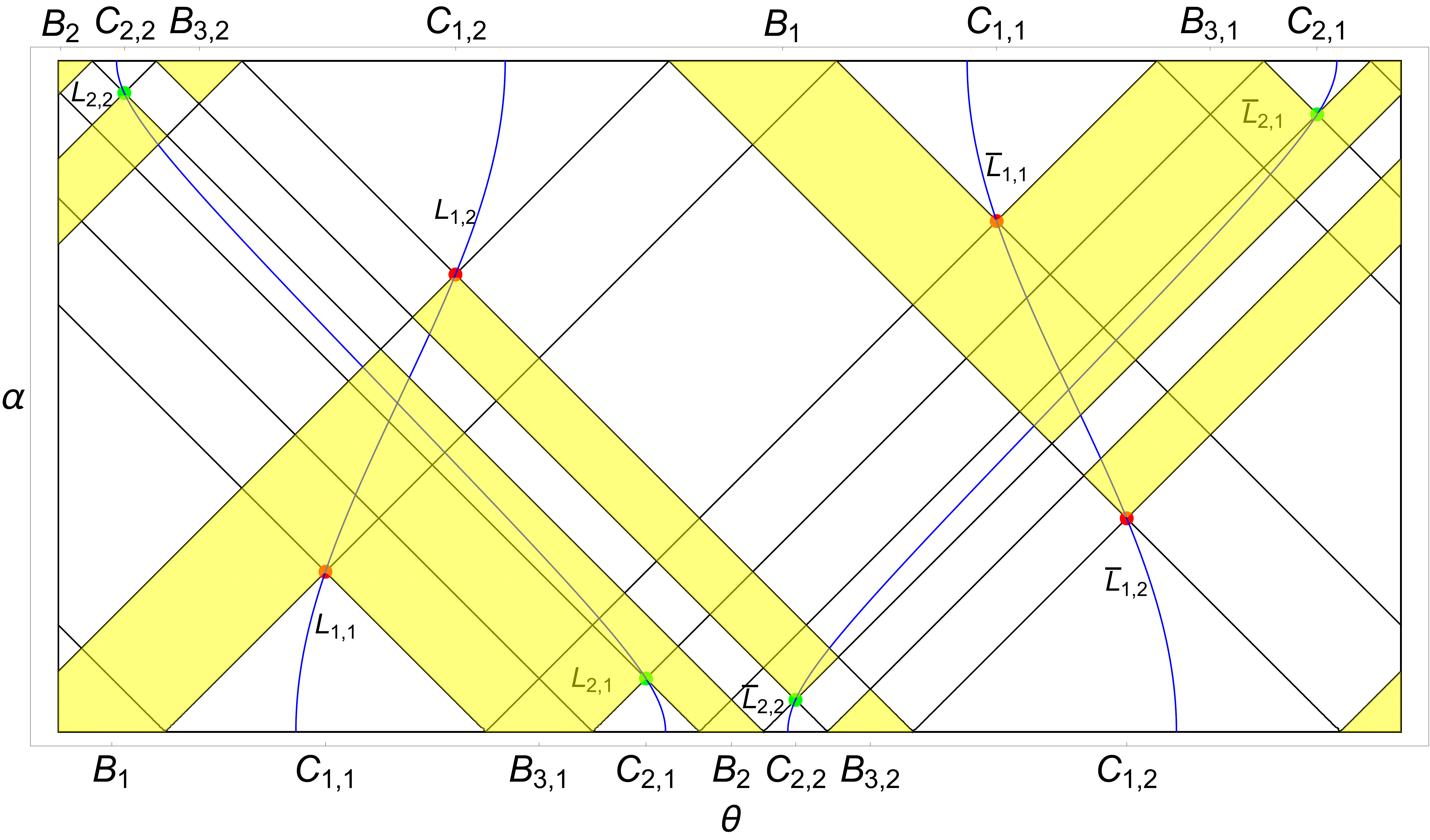}
  \caption{Three-boundary wormhole and its fundamental region in the kinematic space. In the left figure, the two pairs of identified geodesics are shown  in Poincar\'{e} disk, and they divided the boundary into eight regions.
  In the right figure,
 the fundamental region in the kinematic space is shown.
  The red and green points correspond to the identified geodesics,
  and the blue lines are the timelike geodesics formed by normal-geodesics of $\g_1,\g_2$.}
  \label{twofund}
\end{figure}

For a three-boundary wormhole and a torus wormhole, both are defined by the identifications of two pairs of geodesics.
The identifications are generated by  two fundamental elements $\g_1, \g_2$ of the corresponding Fuchsian group $\G=\{\g_1,\g_2\}$.
We denote the four geodesics as $\mathbf{L}_{i,j}$ with $\mathbf{L}_{i,2}=\g_i\mathbf{L}_{i,1}, i=1,2$.
The geodesics and the boundaries in the Poincar\'{e} disk are shown in the left figure of Fig. \ref{twofund}.
In this figure, we choose the identification to get a three-boundary wormhole.
The discussion for other identification is similar.
Now the geodesics on the  Poincar\'{e} disk can be classified into 72 classes, depending on their ending points.
Correspondingly, the kinematic space is separated into 72 regions.

Now let us discuss the fundamental region for this three-boundary wormhole.
Notice that each pair of identified geodesics $\mathbf{L}_{i,1},\mathbf{L}_{i,2}$
can define a BTZ spacetime corresponding to $\g_i$ such that $\mathbf{L}_{i,2}=\g_i\mathbf{L}_{i,1}$,
and we can read  the fundamental region for the resulting BTZ according to the rule we defined above.
Since any points outside this fundamental region can be mapped into it by an element $\g_i^n$,
then the regions including those points must not be a part of the fundamental region for the wormhole.
So the fundamental region of a three-boundary wormhole, as shown in the right figure of Fig. \ref{twofund}, is the intersection
of the fundamental regions of all the BTZ defined by each pair of geodesics.
And this way to choose the fundamental region works for all kinds of multi-boundary wormhole.

\begin{figure}
  \centering
  \includegraphics[width=0.3\linewidth]{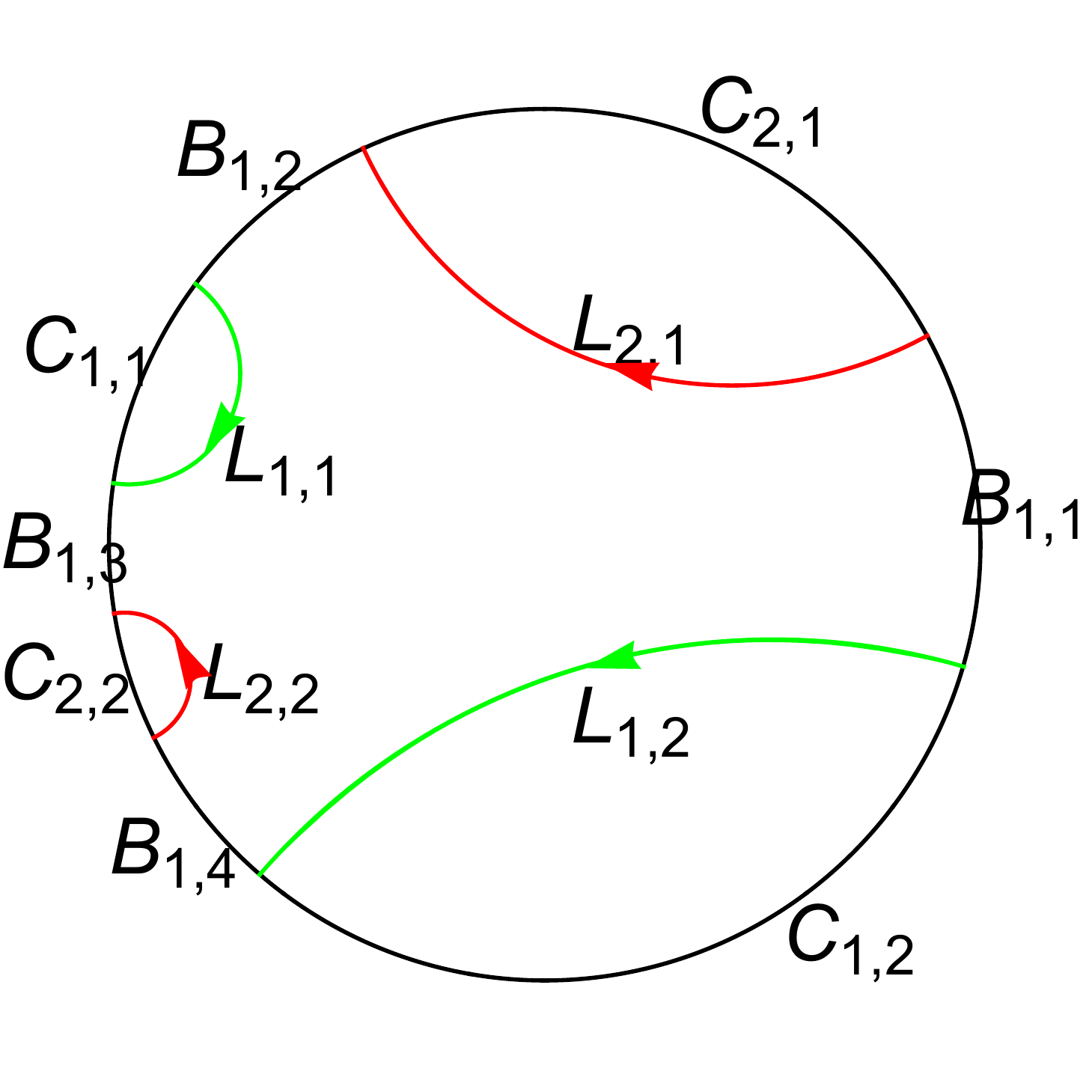}
  \includegraphics[width=0.55\linewidth]{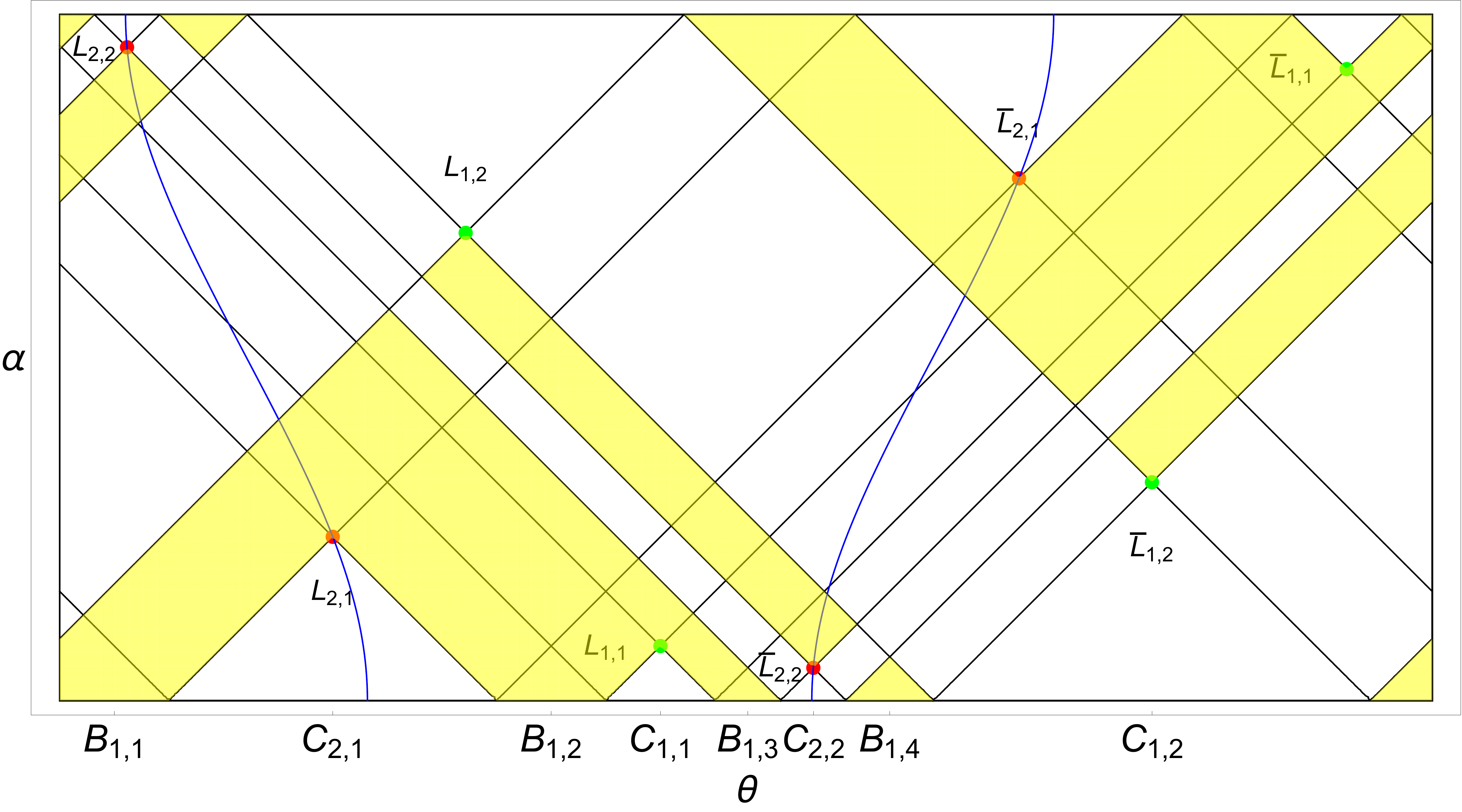}
  \includegraphics[width=0.3\linewidth]{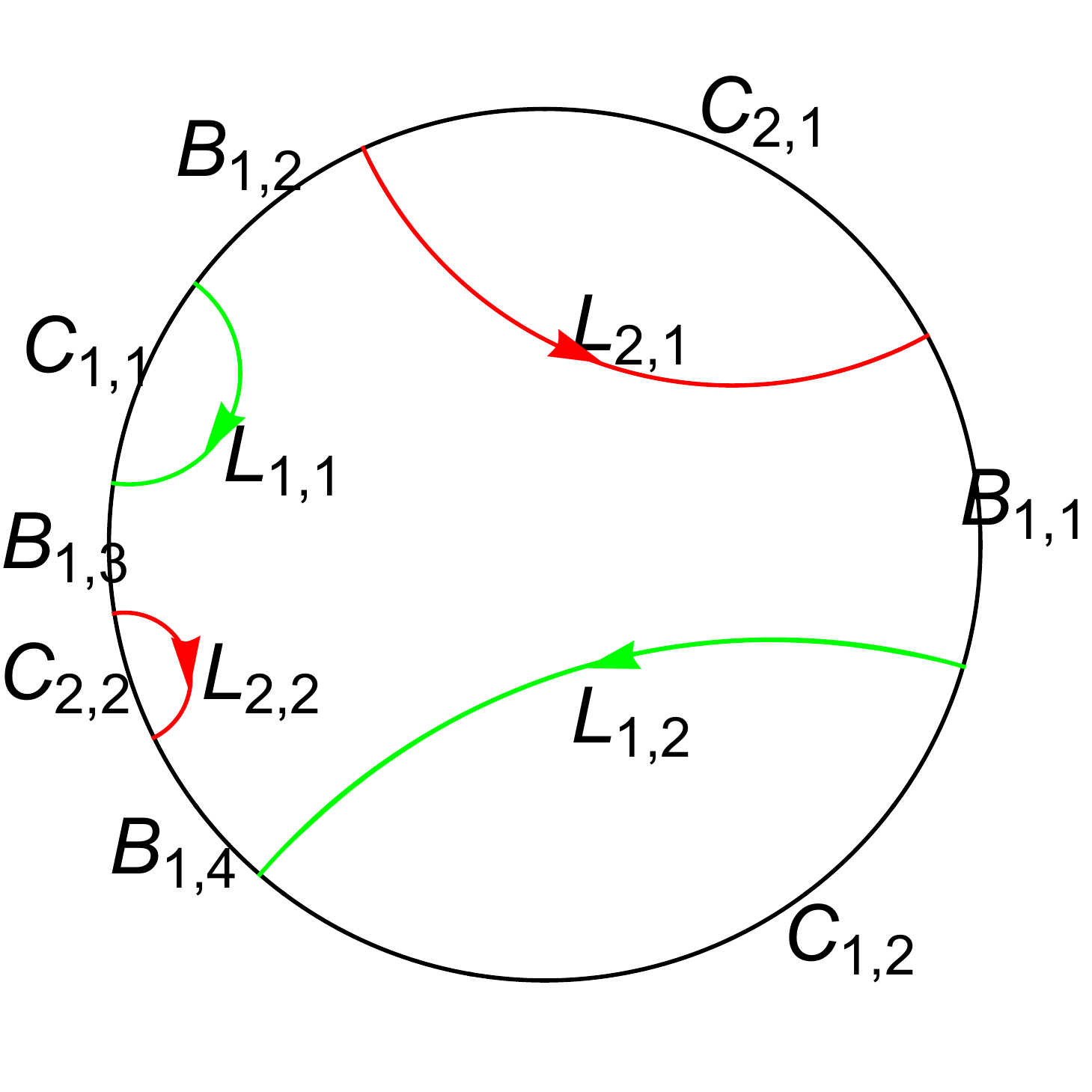}
  \includegraphics[width=0.55\linewidth]{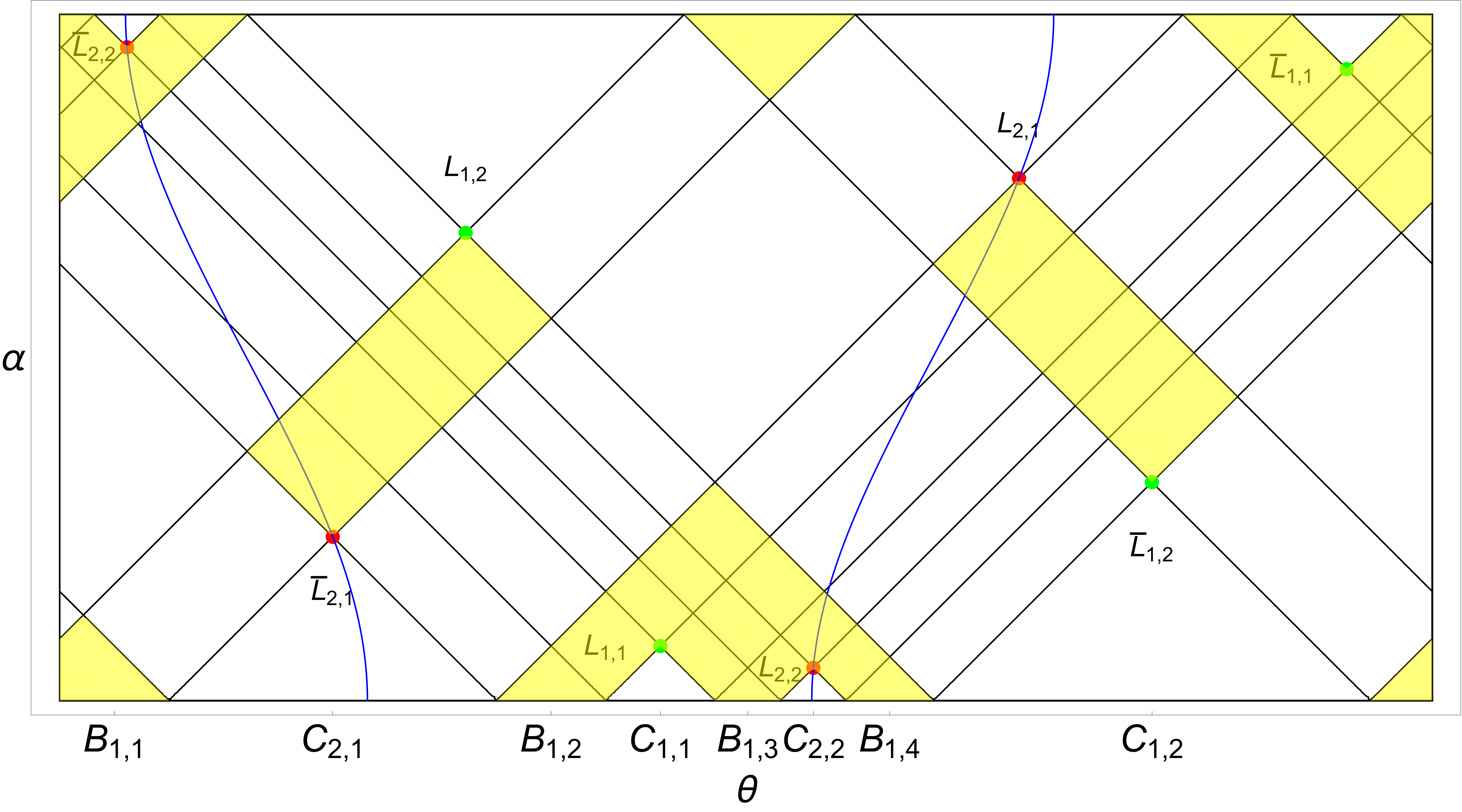}
  \caption{Single-boundary torus wormhole and its fundamental region in the kinematic space for $\G={\g_1,\g_2}$ is shown on the upper half part.
  In the left figure, the two pairs of identified geodesics are shown in Poincar\'{e} disk,
  and they divided the boundary into eight regions.
  In the right figure, the fundamental region in the kinematic space is shown.
  The red and green points correspond to the identified geodesics,
  and the blue lines are the timelike geodesics formed by normal-geodesics of $\g_1,\g_2$.
  On the lower half part,  the same wormhole but the group is generated differently $\G={\g_1,\g_2^{-1}}$.
  In the left figure, the direction of the red geodesics are changed, suggesting the element is $\g_2^{-1}$.
  }
  \label{torfund}
\end{figure}

As we mentioned above, for the same four geodesics  a different kind of identification leads to a single-boundary torus wormhole.
As showed in the upper half of Fig. \ref{torfund}, the geodesics in the same color are identified.
We should notice that the fundamental region in this identification is the same as the three-boundary wormhole in Fig. \ref{twofund}.
This is just because we choose the group to be generated by ${\g_1,\g_2}$.
If we choose the generators to be ${\g_1,\g_2^{-1}}$, the fundamental region is showed in the lower half of Fig. \ref{torfund}.
Although the fundamental region may be the same for different kinds of wormhole,
the same point  corresponds to different kinds of geodesics in these wormholes, since the identification is different.

To read the topology of the fundamental region, one should glue the identified points with same orientation on the boundaries of the fundamental region.
As all the boundaries are parts of the boundary for the fundamental region of some fundamental element $\g_i$,
the way to glue them is just the same as the BTZ case.
But there is some slight difference in the wormhole cases.
For  the three-boundary wormhole,
the naive guess that the topology of its fundamental region in the kinematic space is just two pairs of pants is not correct.
In fact, the rectangle parts in the fundamental region  affects the topology.
The red point representing $L_{1,2}$ should be glued to $L_{1,1}$,
while the green point $\bar{L}_{2,2}$ should be glued to $\bar{L}_{2,1}$.
Then the lower triangle part and the upper triangle part which represents the same geodesics with different orientation
is connected by this rectangle part.
The topology of the fundamental region in the kinematic space turns out to  be a surface with six boundaries. This  can be seen by cutting each pair of pants and gluing them together.
For the torus wormhole, the topology of the fundamental region is a surface of genus 2 with two boundaries.

In order to distinguish different kinds of wormholes, it is not enough to consider only the fundamental region,
which is determined by the fundamental elements in the Fuchsian group.
We need to take the exact identification into account.
One simple way to do this is to draw the geodesics corresponding to the fundamental elements clearly.
For example, the fundamental region in yellow in the upper of Fig. \ref{torfund} looks the same as the one in Fig. \ref{twofund}.
However, the geodesics characterizing the identifications $\g_i$ are obviously different.

\section{Conclusion and discussion}

In this paper, we studied the properties of the kinematic space from geometric points of view.
First of all we showed how the kinematic space of AdS$_3$ can be constructed geometrically in the embedding space.
As every geodesic on the Poincar\'e disk is the boundary of the intersection between the Poincar\'e disk and another disk centered outside,
the kinematic space is actually formed by the tip points of the causal diamond of the other disk in the embedding space.
In this picture, the causal structure in the kinematic space is easily understood.
Moreover we discussed the Fuchsian group and its action on the geodesics to get the multi-boundary wormhole.
We showed that for each $SL(2,R)$ transformation in the Fuchsian group its normal-geodesics make up a geodesic in the kinematic space.
If the transformation is hyperbolic, elliptic or parabolic,
the corresponding geodesic in the kinematic space is timelike, spacelike or null respectively.
More surprisingly, the horizon length of the BTZ wormhole can be read by the length of the corresponding timelike geodesic in the kinematic space.
Finally we discussed the kinematic space for the multi-boundary wormhole.
We started from the kinematic space for global AdS$_3$ and considered the identification of the elements in the Fuchsian group.
For the BTZ blackhole, we defined consistently its fundamental region in the kinematic space.
For the three-boundary wormhole, we argued that its fundamental region in the kinematic space is formed by
the intersection of two fundamental regionals of the BTZ wormhole constructed by two fundamental elements in its Fuchsian group. For the
single-boundary wormhole, its fundamental region could be same as the one for the three-boundary wormhole, but the timelike geodesics
corresponding to the identification are different.

Our study on the kinematic space is purely geometrical, having nothing to do with the differential entropy.
The discussion is quite different from the ones in the literature.
Our approach could be applied to the study of the holographic entanglement entropy and bit threads\cite{Freedman:2016zud}.
We would like to leave them for future study\cite{ZhangChen}.

\vspace*{10mm}
\noindent {\large{\bf Acknowledgments}}\\

The work was in part supported by NSFC Grant No.~11275010, No.~11335012 and No.~11325522.
We would like to thank B. Czech and M. Headrick for helpful discussions.

\end{document}